\newcommand {\imp} {\rightarrow}
\newcommand {\cond} {\Rightarrow}
\newcommand {\orr} {\vee}
\newcommand {\andd} {\land}
\newcommand {\prova} {\vdash}
\newcommand {\nott} {\lnot}

\newcommand {\sx} {\langle}
\newcommand {\dx} {\rangle}
\newcommand {\incluso} {\subseteq}
\newcommand {\appartiene} {\in}
\newcommand {\emme} {\mathcal{M}}
\newcommand {\GI} {\mathcal{G}}
\newcommand {\elle} {\mathcal{L}}

\newcommand {\trans}[1]{\stackrel{#1}{\longrightarrow}}

\newcommand {\unione} {\cup}

\newcommand {\falso} {\bot}
\newcommand {\vero} {\top}

\newcommand {\tc} {\mid}
\newcommand {\vuoto} {\emptyset}
\newcommand {\WW} {\mathcal{W}}
\newcommand {\AAA} {\mathcal{A}}

\newcommand {\B} {\mathcal{B}}
\newcommand {\diverso} {\neq}
\newcommand {\pr} {\circ}

\newcommand {\finedim} {\begin{flushright} $\Box$ \end{flushright}}

\newcommand{\MM}{\mbox{$\mathcal{M}$}}
\newcommand{\BB}{\mbox{$\mathcal{B}$}}

\newcommand{\ff}[1]{\stackrel{#1}{\longrightarrow}}

\newcommand{\be}{\begin{enumerate}}
\newcommand{\ee}{\end{enumerate}}

\newcommand{\hide}[1]{}

\newcommand{\irule}[3]
{\prooftree{#1}\justifies{#2}\using{\:#3}\endprooftree}

\hide{

{\qed \end{trivlist}}

{\qed \end{trivlist}}

\newenvironment{definition}
{\begin{defi} \rm}{\qed \end{defi}}

}

\newcommand {\caL} {{\cal L}}

\newcommand {\alf} {{\cal A}}

\def \cases{\left \{\begin{array}{l}}
\def \endcases{\end{array}\right .}

\newcommand {\ri} {\rightarrow}
\newcommand {\Ri} {\Rightarrow}

\newcommand {\bes} {\begin{description}}
\newcommand{\ens} {\end{description}}
\newcommand {\la} {\langle}
\newcommand {\ra} {\rangle}

\newcommand {\beq} {\begin{quote}}
\newcommand {\enq} {\end{quote}}
\newcommand {\bit} {\begin{itemize}}
\newcommand {\enit} {\end{itemize}}

\documentclass[acmtocl]{acmtrans2m}
\usepackage{prooftree}
\usepackage[dvips]{graphicx}
\usepackage{latexsym}
\usepackage{amssymb}
\usepackage{color}

\title{A Sequent Calculus and a Theorem Prover for Standard Conditional Logics}
\author{NICOLA OLIVETTI and
GIAN LUCA POZZATO \\ Universit\`a di Torino \and CAMILLA B.
SCHWIND \\ \'Ecole d'Architecture de Marseille}
\begin{abstract}
  In this paper we present a cut-free sequent calculus, called SeqS, for some standard conditional
  logics, namely CK, CK+ID, CK+MP and CK+MP+ID. The calculus uses labels and transition formulas
  and can be used to prove decidability and
  space complexity bounds for the respective logics. We also present
  CondLean, a theorem prover for these logics implementing SeqS
  calculi written in SICStus Prolog.

\end{abstract}

 \terms{Languages,
Logic, Theory} 

\newtheorem{teorema}{Theorem}[section]
\newtheorem{lemmaPosu}[teorema]{Lemma}
\newtheorem{corollario}[teorema]{Corollary}
\newtheorem{proposizione}[teorema]{Proposition}
\newtheorem{defi}[teorema]{Definition}
\newtheorem{exa}[teorema]{Example}
\newtheorem{rem}[teorema]{Remark}

\newtheorem{definition}[teorema]{Definition}

\begin{document}

\maketitle

\begin{bottomstuff}
Authors' addresses: N. Olivetti, Dipartimento di Informatica -
Universit\`a degli Studi di Torino, corso Svizzera 185 - 10149
Turin - Italy, e-mail: {\tt olivetti@di.unito.it}.\\ G.L. Pozzato,
Dipartimento di Informatica - Universit\`a degli Studi di Torino,
corso Svizzera 185 - 10149 Turin - Italy, e-mail: {\tt
pozzato@di.unito.it}.\\ C.B. Schwind, \'Ecole d'Architecture de
Marseille - Luminy, 184 avenue de Luminy - 13288 Marseille cedex 9
- France, \\ e-mail: {\tt Camilla.Schwind@map.archi.fr}
\end{bottomstuff}

\section{Introduction}
 Conditional logics have a long history. They have  been studied first by Lewis
(\cite{Lewis:73,Nute80,Chellas,Stalnaker}) in order to formalize a
kind of hypothetical reasoning (if $A$ were the case then $B$),
that cannot be captured by
 classical  logic with  material implication.\\

In the last years, interesting applications of
 conditional logic to several domains of  artificial intelligence such as
 knowledge representation, non-monotonic reasoning,
belief revision,
 representation of counterfactual sentences, deductive
 databases have been proposed  (\cite{crocco2}). For instance,
 in \cite{Grahne:91JLC}     knowledge and database update is
 formalized by some conditional logic.   Conditional logics have also been
used to modelize belief revision
(\cite{belief-revision,ramsey-test,GiordanoGliozziOlivetti:98,16}).
Conditional logics can provide an axiomatic foundation of
non-monotonic reasoning (\cite{KrausLehmannMagidor:90}), as it
turns out that all forms of inference studied in the framework of
non-monotonic (preferential) logics are particular cases of
conditional axioms (\cite{CroccoLamarre:92}). Causal inference,
which is very important for applications in action planning
(\cite{Schwind:99}), has been modelled by conditional logics
(\cite{GiordanoSchwind:04}). Conditional Logics have been used to
model hypothetical queries in deductive databases and logic
programming; the conditional logic CK+ID is the basis of the logic
programming language defined in \cite{12}. In system diagnosis,
conditional logics  can be used to reason hypothetically about
 the expected  functioning of system components with respect to the
observed  faults.  \cite{diagnosi} introduces a conditional logic,
DL, suitable for diagnostic reasoning and which allows to
represent and reason with assumptions in model-based diagnosis.
Another interesting application of conditional logics is the
formalization of  \emph{prototypical reasoning}, that is to say
reasoning about typical properties and exceptions. Delgrande
in \cite{Delgrande:87} proposes  a conditional logic for
prototypical reasoning.

Finally, an obvious application concerns natural language
semantics where conditional logics are used
 in order to give a formal treatment of hypothetical and
counterfactual sentences as presented in \cite{Nute80}. A broader
discussion about counterfactuals can be found in
\cite{counterfactuals}.

In spite of their significance,  very few  proof systems have been
proposed  for conditional logics: we just mention
\cite{Lamarre:94,GroeneboerDelgrande:90,CroccoFarinas:95,Governatori:02,Gent,7,GGOSTableaux2003}.
One possible reason of the underdevelopment of proof-methods for
conditional logics is  the lack of a universally accepted
semantics for them. This is in sharp contrast to modal and
temporal logics which  have a consolidated semantics based on a
standard kind of Kripke structures.

Similarly to modal logics, the semantics of conditional logics can
be defined in terms of possible world structures. In this respect,
conditional logics can  be seen as a generalization of modal
logics (or a type of multi-modal logic) where the conditional
operator is a sort of  modality indexed by a formula of the same
language.\\The two  most  popular semantics for conditional logics
are  the so-called {\em sphere semantics} (\cite{Lewis:73}) and
the {\em selection function semantics} (\cite{Nute80}). Both are
possible-world semantics, but are based on  different (though
related) algebraic notions. Here we adopt the selection function
semantics, which is more general than the sphere semantics.

Since we adopt the selection function semantics, {\bf CK} is the
fundamental system; it has the same role as the system K (from
which it derives its name) in modal logic: CK-valid formulas are
formulas that are valid in every selection function model.

In this work we present a sequent calculus for CK and for three
standard extensions of it, namely CK+ID, CK+MP\footnote{This
conditional system is related to modal logic T.} and CK+MP+ID.
This calculus makes use of labels, following the line of
\cite{vigano} and \cite{dov}. To the best of our knowledge, this
is the first calculus for these systems. Some tableaux calculi
were developed in \cite{GGOSTableaux2003} and in
\cite{OlivettiSchwind:99} for other more specific conditional
systems.

Our goal is to obtain a decision procedure for the logics under
consideration. For this reason, we undertake a proof theoretical
analysis of our calculi. In order to get a terminating calculus,
it is crucial to control the application of the \emph{contraction
rule}. This rule allows for duplicating a formula in a backward
proof search and thus is a potential source of an infinite
expansion of a branch. Generally speaking, the status of the
contraction rules varies for different logical systems: in some
cases, contraction rules can be just removed without losing
completeness, in some others they cannot be eliminated, but their
application can be controlled in such a way that the branch
expansion terminates. This is what happens also for our
conditional logics. In other cases, contraction rules can be
eliminated, but at the price of changing the logical rules, as it
happens in \cite{Hudelmaier}.

In this work, we show that the contraction rules can be eliminated
in the calculi for CK and CK+ID. In this way, the calculus not
only provides a decision procedure, but it can also be used to
establish a complexity bound for these logics (the decidability
for these logics has been shown in \cite{Nute80}). Roughly
speaking, if the rules are analytic, the length of each branch is
bounded essentially by the length of the initial sequent;
therefore, we can easily obtain that the calculus give a
polynomial space complexity.

For CK+MP and CK+MP+ID the situation is different: contraction
rules cannot be eliminated without losing completeness. However,
we show that they can be used in a controlled way, namely it is
necessary to apply the contraction at most one time on each
conditional formula of the form $x: A \cond B$ in every branch of
a proof tree. This is sufficient to obtain a decision procedure
for these logics.

It is worth noting that the elimination of contractions is
connected with a remarkable property, the so-called
\emph{disjunction property} for conditional formulas: if ($A_1
\cond B_1$) $\orr$ ($A_2 \cond B_2$) is valid, then either ($A_1
\cond B_1$) or ($A_2 \cond B_2$) is valid too.

As a difference with modal logics, for which there are lots of
efficient implementations (\cite{leanTAP}, \cite{leanTAP-Rev},
\cite{free-variable-tableau}), to the best of our knowledge very
few theorem provers have been implemented for conditional logics
(\cite{Lamarre:94} and \cite{Governatori:02}). We present here a
simple implementation of our sequent calculi, called {\bf
CondLean}; it is a Prolog program which follows the \emph{lean}
methodology (\cite{leanTAP}, \cite{leanTAP-Rev}), in which every
clause of a predicate \texttt{prove} implements an axiom or rule
of the calculus and the proof search is provided for free by the
mere depth-first search mechanism of Prolog, without any ad hoc
mechanism. We also present an alternative version of our theorem
prover inspired by the tableau calculi for modal logics introduced
in \cite{free-variable-tableau}.

The plan of the paper is as follows: in section 2 we introduce the
conditional systems we consider, in section 3 we present the
sequent calculi for conditional systems above. In section 4 we
analyze the calculi in order to obtain a decision procedure for
the basic conditional system, CK, and for the three mentioned
extensions of it. In section 5 we present the theorem prover
CondLean. In section 6 we discuss some related work.

\section{Conditional Logics}
Conditional logics are extensions of classical logic obtained by
adding the conditional operator $\cond$. In this paper, we only
consider propositional conditional logics.

A propositional conditional language $\elle$ contains the
following items:
\begin{list}{-}{}
  \item a set of propositional variables \emph{ATM};
  \item the symbol of \emph{false} $\falso$;
  \item a set of connectives\footnote{The usual connectives $\vero$, $\andd$, $\orr$ and $\nott$
    can be defined in terms of $\falso$ and $\imp$.} $\imp$, $\cond$.
\end{list}
We define formulas of $\elle$ as follows:
\begin{list}{-}{}
  \item $\falso$ and the propositional variables of \emph{ATM} are
  \emph{atomic formulas};
  \item if \emph{A} and \emph{B} are formulas, $A$ $\imp$ $B$ and
  $A$ $\cond$ $B$ are \emph{complex formulas}.
\end{list}

We adopt the \emph{selection function semantics}. We consider a
non-empty set of possible worlds $\WW$. Intuitively, the selection
function $f$ selects, for a world $w$ and a formula $A$, the set
of worlds of $\WW$ which are \emph{closer} to $w$ given the
information $A$. A conditional formula $A \cond B$ holds in a
world $w$ if the formula $B$ holds in \emph{all the worlds
selected by $f$ for $w$ and $A$}.

A model is a triple:
\begin{center}
  $\emme$ = $\sx$ $\WW$, $f$, [ ] $\dx$
\end{center}
where:
\begin{list}{-}{}
  \item $\WW$ is a non empty set of items called \emph{worlds};
  \item $f$ is the so-called \emph{selection function} and has the
  following type:
  \begin{center}
    $f$: $\WW$ $\times$ $2^{\WW}$ $\longrightarrow$ $2^{\WW}$
  \end{center}
  \item $[$ $]$ is the \emph{evaluation function}, which assigns to an
  atom $P$ $\appartiene$ $ATM$ the set of worlds where $P$ is
  true, and is extended to the other formulas as follows:
    \begin{list}{}{}
      \item * [$\falso$] = $\vuoto$
      \item * [$A$ $\imp$ $B$]=($\WW$ - [$A$]) $\unione$ [$B$]
      \item * [$A$ $\cond$ $B$]=\{$w$ $\appartiene$ $\WW$ $\tc$ $f$($w$, [$A$]) $\incluso$ [$B$]\}
    \end{list}
\end{list}

Observe that we have defined $f$ taking [$A$] rather than $A$
(i.e. $f$($w$,[$A$]) rather than $f$($w$,$A$)) as argument; this
is equivalent to define $f$ on formulas, i.e. $f$($w$,$A$) but
imposing that if [$A$]=[$A^{'}$] in the model, then $f$($w,
A$)=$f$($w, A^{'}$). This condition is called \emph{normality}.

The semantics above characterizes the \emph{basic conditional
system}, called {\bf CK}. An axiomatization of the CK system is
given by:
\begin{list}{-}{}
\item all tautologies of
classical propositional logic.

\vspace{0.2cm}

\item (Modus Ponens) \quad $\irule{A \quad A \ri B}{B}{}$

\vspace{0.2cm}

\item (RCEA)  \quad
$\irule{A \leftrightarrow B} {(A\Ri C)\leftrightarrow(B\Ri C)} {}$

\vspace{0.2cm}

\item (RCK) \quad
$\irule{(A_1 \wedge \dots \wedge A_n ) \rightarrow B}%
{(C\Ri A_1 \wedge \dots \wedge C\Ri A_n)\rightarrow(C\Ri B)} {}$
\end{list}
Other conditional systems are obtained by assuming further
properties on the selection function; we consider the following
three standard extensions of the basic system CK:

\begin{center} \(\begin{array}{|c|c|c|} \hline
 \mbox{System} & \mbox{Axioms}  &  \mbox{Model condition} \\
\hline
{\bf CK+ID} & A \Ri A  &  f(x,[A]) \subseteq [A]  \\
\hline {\bf CK+MP} &  (A \Ri B)   \ri ( A \rightarrow  B) &
w \in [A]  \ri  w\in f(w,[A])  \\
\hline {\bf CK+MP+ID} &  (A \Ri B)   \ri ( A \rightarrow  B), & w
\in [A] \ri w\in f(w,[A]),  \\
     &  A \Ri A  &  f(x,[A]) \subseteq [A]
\\\hline
\end{array}\)
\end{center}

\section{A Sequent Calculus for Conditional Logics}
In this section we present {\bf SeqS}, a sequent calculs for the
conditional systems introduced above. S stands for \{CK, ID, MP,
ID+MP\}; the calculi make use  of labels to represent possible
worlds.\\We consider a conditional language $\elle$ and a
denumerable alphabet of labels $\AAA$, whose elements are denoted
by \emph{x, y, z, ...}.\\There are two kinds of formulas:
\begin{enumerate}
\item \emph{labelled formulas}, denoted by \emph{x: A}, where
  \emph{x $\appartiene$ $\AAA$} and $A\in \cal L$, used to represent that \emph{A} holds in a world \emph{x};
\item  \emph{transition formulas},  denoted by \emph{x $\trans{A}$
  y}, where \emph{x, y
  $\appartiene$ $\AAA$} and $A\in \cal L$.  A transition
  formula \emph{x $\trans{A}$ y} represents that \emph{y $\appartiene$ f}(\emph{x},
  [\emph{A}]).
\end{enumerate}
A {\bf sequent} is a pair $\sx \Gamma, \Delta \dx$, usually
denoted with $\Gamma\prova\Delta$, where $\Gamma$ and $\Delta$ are
multisets of formulas. The intuitive meaning of
$\Gamma\prova\Delta$ is: every model that satisfies all labelled
formulas  of $\Gamma$ in the respective worlds (specified by the
labels) satisfies at least one of the labelled formulas of
$\Delta$ (in those worlds). This is made precise by the notion of
{\em validity}  of a sequent given in the next definition:

\begin{definition}[Sequent validity]
Given a model
\begin{center}
$\MM=\la\WW$\emph{, f, }$[~]\ra$
\end{center}
for $\caL$, and a label alphabet $\alf$, we consider any {\em
mapping}
\begin{center}
  $I: \alf \to \WW$
\end{center}
Let $F$ be a labelled formula, we define $\MM\models_I F$ as
follows:
\begin{list}{-}{}
  \item $\MM\models_I$ $x$: $A$ \emph{iff} $I(x)$ $\in$
$[A]$
  \item $\MM\models_I$ $x$ $\ff{A}$ $y$ \emph{iff} $I(y)$ $\in$
$f(I(x)$, $[A])$
\end{list}
 We say that $\Gamma \vdash \Delta$ is {\em
valid} in $\MM$ if for every mapping $I: \alf \to \WW$, if $\MM
\models_I F$ for every $F\in\Gamma$, then $\MM\models_I G$ for
some $G\in\Delta$. We say that $\Gamma \vdash \Delta$ is valid in
a system
 (CK or one of its extensions) if it is valid in every
$\MM$ satisfying the specific conditions for that system (if any).
\end{definition}

In Figure \ref{Figura con SeqS} we present the calculi for CK and
its mentioned extensions.

\begin{figure}
\begin{scriptsize}
\fbox{ \(
\begin{array}{rl@{\quad\quad}rl}
{\bf (AX)} & \Gamma,F \vdash \Delta,F & {\bf (A\bot) } & \Gamma,
x: \bot \vdash \Delta
\\
\\
{\bf (Weak L)} & \irule{\Gamma \vdash \Delta} {\Gamma,F \vdash
\Delta} {} & {\bf (Weak R)} & \irule{\Gamma \vdash \Delta} {\Gamma
\vdash \Delta, F} {}
\\
\\
{\bf (Contr L)} & \irule{\Gamma, F,F \vdash \Delta} {\Gamma,F
\vdash \Delta} {} & {\bf (Contr R)} & \irule{\Gamma \vdash
\Delta,F,F} {\Gamma \vdash \Delta,F} {}
\\
\\
{\bf (\ri R)} & \irule{\Gamma, x: A \vdash x: B,\Delta} {\Gamma
\vdash x: A \ri B,\Delta} {} & {\bf (\ri L)} & \irule{\Gamma
\vdash x: A,\Delta\quad \Gamma, x: B \vdash \Delta} {\Gamma,  x: A
\ri B \vdash \Delta} {}
\\
\\
\\
{\bf (EQ)} &
\irule{u: A \vdash u:B \quad\quad  u:B \vdash u:A}%
{\Gamma, x\ff{A} y \vdash  x\ff{B} y, \Delta }%
{} & &
\\
\\
{\bf (\Ri L)} & \irule{\Gamma\vdash x\ff{A} y,  \Delta  \quad\quad
\Gamma, y:B \vdash \Delta }%
{ \Gamma, x: A \Ri B \vdash \Delta  }%
{} & {\bf (\Ri R)} &
\irule{\Gamma,  x\ff{A} y \vdash y: B, \Delta}%
{\Gamma\vdash  x: A \Ri B, \Delta  }%
{(y \not\in \Gamma, \Delta)}
\\
\\
\\
{\bf (ID)} &
\irule{\Gamma, y:A\vdash\Delta }%
{ \Gamma,  x\ff{A} y \vdash \Delta  }%
{} & {\bf (MP)} & \irule{\Gamma \vdash x: A, \Delta} {\Gamma\vdash
x\ff{A} x, \Delta  } {}
\\
\\
\end{array}
\) } \end{scriptsize}\caption{Sequent calculi SeqS; the (ID) rule
is for SeqID and SeqID+MP only; the (MP) rule is for SeqMP and
SeqID+MP only.} \label{Figura con SeqS}
\end{figure}

\begin{figure}

\fbox{ \(
\begin{array}{rl@{\quad}rl}
{\bf (\andd L)} & \irule{\Gamma, x: A, x: B \vdash \Delta}
{\Gamma, x: A \andd B \vdash \Delta} {} & {\bf (\andd R)} &
\irule{\Gamma \vdash \Delta, x: A \quad\quad \Gamma \vdash \Delta,
x: B} {\Gamma \vdash \Delta, x: A \andd B} {}
\\
\\
{\bf (\orr L)} & \irule{\Gamma, x: A \vdash \Delta \quad\quad
\Gamma, x: B \vdash \Delta} {\Gamma, x: A \orr B \vdash \Delta} {}
& {\bf (\orr R)} & \irule{\Gamma \vdash \Delta, x: A, x: B}
{\Gamma \vdash \Delta, x: A \orr B} {}
\\
\\
{\bf (\nott L)} & \irule{\Gamma \vdash \Delta, x: A} {\Gamma, x:
\nott A \vdash \Delta} {} & {\bf (\nott R)} & \irule{\Gamma, x: A
\prova \Delta} {\Gamma \vdash \Delta, x: \nott A} {}
\\
\\
\\
{\bf (A\vero)} & \Gamma \prova \Delta, x: \vero
\\
\\
\end{array}
\) } \caption{Additional axioms and rules in SeqS for the other
boolean operators, derived from the rules in Figure \ref{Figura
con SeqS} by the usual equivalences.} \label{Regole aggiuntive di
SeqS}
\end{figure}

\begin{exa}\label{ex1}
We show a derivation of the (ID) axiom.
\[\begin{prooftree}
\[y: A \vdash y: A
\justifies x \ff{A} y \vdash y: A \using (ID)
\]
\justifies \vdash x: A \Ri A \using (\Ri R)
\end{prooftree}\]
\end{exa}

\begin{exa}\label{ex2}
We show a derivation of the (MP) axiom.
\[\begin{prooftree}
\[
  \[
    \[
    x: A \vdash x: A, x: B
    \justifies
    x: A \vdash x\ff{A} x, x: B
      \using (MP)
    \]
    x: A, x: B \vdash x: B
    \justifies
    x: A \Ri B, x: A\vdash x: B
    \using (\Ri L)
 \]
    \justifies
    x: A \Ri B   \vdash x: A\ri B
    \using (\imp R)
\]
    \justifies
    \vdash x:  (A \Ri B) \ri (A\ri B)
    \using (\imp R)
\end{prooftree}
\]
\end{exa}

In the following, we will need to consider the
\emph{permutability} of a rule over another one\footnote{In
general, we say that a rule $r_1$ permutes over a rule $r_2$ if
the following condition holds: if they are both applicable to a
sequent $\Gamma \prova \Delta$ and there is a proof tree where
$r_2$ is applied to $\Gamma$ $\prova$ $\Delta$ and $r_1$ is
applied to one of the premises of $r_2$, then there exists a proof
tree of $\Gamma$ $\prova$ $\Delta$ where $r_1$ is applied to the
sequent $\Gamma$ $\prova$ $\Delta$ and $r_2$ is applied to one of
the premises of $r_1$.}. It is easy to observe the following
Lemma:

\begin{lemmaPosu}[Permutability of the rules]
All the SeqS's rules permute over the other rules, with the
exception of ($\cond$ L), which does not always permute over
($\cond$ R).
\end{lemmaPosu}
In particular, it does not permute over ($\cond$ R) which
introduces the label \emph{y} used to apply ($\cond$ L), as shown
in the following counterexample in SeqCK:

\[
\begin{prooftree}
\[
  x \trans{A} y \prova x \trans{A} y, y: B
  \[
    y: B, y: C, x \trans{A} y \prova y: B
    \justifies y: B \andd C, x \trans{A} y \prova y: B \using
    (\andd L)
  \]
  \justifies x: A \cond (B \andd C), x \trans{A} y \prova y: B \using (\cond L)
\]
\justifies x: A \cond (B \andd C) \prova x: A \cond B  \using
(\cond R)
\end{prooftree}
\]

\noindent The application of ($\cond$ L) with \emph{x}, the only
available label, leads to a proof which fails:

\[
  \begin{prooftree}
    \[
     x \trans{A} y \prova y: B, x \trans{A} x
    \justifies \prova x: A \cond B, x \trans{A} x \using (\cond R)
    \]
    \[
      \[
        x: B, x: C, x \trans{A} y \prova y: B
        \justifies x: B, x: C \prova x: A \cond B \using (\cond R)
      \]
      \justifies x: B \andd C \prova x: A \cond B \using (\andd L)
    \]
\justifies x: A \cond (B \andd C) \prova x: A \cond B  \using
(\cond L)
  \end{prooftree}
\]

\noindent If $A, B$ and $C$ are atoms, neither of the two branches
can be closed.\\

The sequent calculus SeqS is sound and complete with respect to
the semantics.

\begin{teorema}[Soundness]\label{soundness}
If $\Gamma \vdash \Delta$ is derivable in SeqS then it is valid in
the corresponding system.
\end{teorema}
\noindent \emph{Proof}. By induction on the height of a derivation
of $\Gamma \vdash \Delta$. As an example, we examine the cases of
($\Ri$ R) and (MP). The other cases are left to the reader.
\begin{list}{-}{}
\item ($\Ri$ R) Let $\Gamma\vdash \Delta, x: A\Ri B$ be derived from
(1)~$\Gamma,x\ff{A} y\vdash \Delta, y:B$, where $y$ does not occur
in $\Gamma$, $\Delta$ and it is different from $x$. By induction
hypothesis we know that the latter sequent is valid. Suppose the
former  is not, and that it is not valid in a model $\MM=\la
\WW,f,[~]\ra$, via a mapping $I$, so that we have: \beq $\MM
\models_I  F$ for every $F\in\Gamma$, $\MM \not\models_I F$ for
any $F\in\Delta$ and $M\not\models_I x: A\Ri B$. \enq As
$M\not\models_I x: A\Ri B$  there exists $w\in f(I(x), [A])-[B]$.
We can define an interpretation $I'(z) = I(z)$ for $z\not=y$ and
$I'(y) = w$. Since $y$ does not occur in $\Gamma$, $\Delta$ and is
different from $x$, we have that $\MM \models_{I'}   F$ for every
$F\in\Gamma$, $\MM \not\models_{I'} F$ for any $F\in\Delta$,
$\MM\not\models_{I'} y: B$ and $\MM \models_{I'} x\ff{A} y$,
against the validity of (1).

\item (MP) Let  $\Gamma\vdash \Delta, x\ff{A} x$ be derived from (2)
$\Gamma\vdash \Delta, x:A$. Let (2) be valid and let $\MM=\la
\WW,f,[~]\ra$ be a model satisfying the MP condition. Suppose that
for one mapping $I$, $\MM \models_I  F$ for every $F\in\Gamma$,
then by the validity of (2) either $\MM\models_I G$ for some $G\in
\Delta$, or $\MM\models_I x: A$. In the latter case, we have $I(x)
\in [A]$, thus $I(x)\in f(I(x),[A])$, by MP, this means that
$\MM\models_I x\ff{A} x$.
\end{list}
\finedim

Completeness is an easy consequence of the admissibility of cut.
By cut we mean the following rule:
\[\irule{\Gamma\vdash \Delta, F \quad  F, \Gamma \vdash \Delta}
{\Gamma \vdash \Delta} {(cut)}\] where $F$ is any labelled
formula. To prove cut adimissibility, we need the following lemma
about label substitution.

\begin{lemmaPosu}\label{lemma label substitution}
If a sequent $\Gamma \vdash \Delta$ has a derivation of height
$h$, then $\Gamma[x/y] \vdash \Delta[x/y]$ has a derivation of
height $h$, where $\Gamma[x/y] \vdash \Delta[x/y]$ is the sequent
obtained from $\Gamma \vdash \Delta$ by replacing a label $x$ by a
label $y$ wherever it occurs.
\end{lemmaPosu}

\noindent \emph{Proof}. By a straightforward induction on the
height of a derivation. \finedim

\begin{teorema}[Admissibility of cut]\label{cut}
If $\Gamma \vdash \Delta, F$ and $F, \Gamma\vdash \Delta$ are
derivable, so $\Gamma \vdash \Delta$.
\end{teorema}

\noindent \emph{Proof}. As usual, the proof proceeds by a double
induction over the complexity of the cut formula and the sum of
the heights of the derivations of the two premises of the cut
inference, in the sense that we replace one cut by one or several
cuts on formulas of smaller complexity, or on sequents derived by
shorter derivations. We have several cases: $(i)$ one of the two
premises is an axiom, $(ii)$ the last step of {\em one} of the two
premises is obtained by a rule in which $F$ is {\em not} the
principal formula\footnote{The principal formula of an inference
step is the formula introduced by the rule applied in that step.},
$(iii)$ $F$ is the principal formula in the last step of {\em
both} derivations.
\begin{description}
\item[(i)] If one of the two premises is an axiom then either $\Gamma \vdash
\Delta$ is an axiom, or the premise which is not an axiom contains
two copies of $F$ and $\Gamma \vdash \Delta$ can be obtained by
contraction.
\item[(ii)]
We distinguish two cases: the sequent where $F$ is not principal
is derived by any rule (R), except the (EQ) rule. This case is
standard, we can permute (R) over the cut: i.e. we cut the
premise(s) of (R) and then we apply (R) to the result of cut. If
one of the sequents, say $\Gamma\vdash\Delta,F$ is obtained by the
(EQ) rule, where $F$ is not principal, then also
$\Gamma\vdash\Delta$ is derivable by the (EQ) rule and we are
done.
\item[(iii)]
$F$ is the principal formula in both the inferences steps leading
to the two cut premises. There are six subcases: $F$ is introduced
by $(a)$ a classical rule, $(b)$ by $(\Ri L), (\Ri R)$, $(c)$ by
(EQ), $(d)$ $F$ by (ID) on the left and by (EQ) on the right,
$(e)$ by (MP) on the left and by (EQ) on the right, $(f)$  by (ID)
on the left and by (MP) on the right. The list is exhaustive. \bes
\item[(a)]
This case is standard and left to the reader.

\item[(b)]
$F= x:A\Ri B$ is introduced by $(\Ri R)$ and $(\Ri L)$. Then we
have
\[
\begin{prooftree}
    \[
    (*) \,\Gamma, x \ff{A} z \vdash z: B,\Delta
    \justifies
    \Gamma \vdash x: A\Ri B,\Delta
    \using
    (\Ri R)
    \]
    \[
    \Gamma \vdash x \ff{A} y, \Delta \quad \Gamma, y: B \vdash \Delta
    \justifies
    \Gamma, x: A\Ri B \vdash \Delta
    \using (\Ri L)
    \]
\justifies \Gamma \vdash \Delta \using (cut)
\end{prooftree}
\]
where $z$ does not occur in $\Gamma,\Delta$ and $z\neq x$; By
Lemma \ref{lemma label substitution}, we obtain that $\Gamma, x
\ff{A} y \vdash y: B,\Delta$ is derivable by a derivation of no
greater height than (*); thus  we  can replace the cut as follows
\[\begin{prooftree}
\[
    \[ \Gamma \vdash x\ff{A} y, \Delta
    \justifies
\Gamma \vdash x\ff{A} y, \Delta, y: B \using (WeakR)
    \]
    \Gamma, x \ff{A} y \vdash y: B,\Delta
    \justifies
\Gamma \vdash \Delta, y:B \using (cut)
\]
\Gamma, y:B \vdash \Delta \justifies \Gamma \vdash \Delta \using
(cut)
\end{prooftree}
\]
The upper cut uses the induction hypothesis on the height, the
lower the induction hypothesis on the complexity of the formula.

\item[(c)]
$F= x\ff{B} y$ is introduced by (EQ) in both premises, we have
\[
\begin{prooftree}
\[
(5)\,u: A \vdash u: B \quad (6)\,u: B \vdash u: A \justifies
\Gamma', x \ff{A} y \vdash x \ff{B} y, \Delta \using  (EQ)
\]
\[
(7)\,u: B \vdash u: C \quad (8)\,u: C \vdash u: B \justifies
\Gamma, x \ff{B} y \vdash x \ff{C} y, \Delta' \using (EQ)
\]
\justifies \Gamma', x \ff{A} y \vdash x \ff{C} y, \Delta' \using
(cut)
\end{prooftree}
\]
where $\Gamma = \Gamma', x \ff{A} y$, $\Delta = x \ff{C} y,
\Delta'$.
 (5)-(8) have been derived by a shorter derivation; thus we can replace the cut by cutting (5)
and (7) on the one hand, and (8) and (6) on the other,  which give
respectively
\begin{center}
(9) $u: A \vdash u: C$ and (10) $u: C \vdash u: A$.
\end{center}
Using (EQ) we obtain $\Gamma', x \ff{A} y \vdash \Delta', x \ff{C}
y$
\item[(d)]
$F= x\ff{B} y$ is introduced on the left by (ID) rule, and it is
introduced on the right by (EQ). Thus we have
\[
\begin{prooftree}
\[u: A \vdash u: B \quad u: B \vdash u: A
\justifies \Gamma', x\ff{A} y\vdash \Delta, x \ff{B} y \using (EQ)
\]
\[ \Gamma', x\ff{A} y, y: B \vdash \Delta
\justifies x\ff{B} y, \Gamma', x\ff{A} y \vdash \Delta \using \rm
(ID)
\]
\justifies \Gamma', x\ff{A} y\vdash \Delta \using \rm (cut)
\end{prooftree}
\]
where $\Gamma = \Gamma', x\ff{A} y$. By Lemma \ref{lemma label
substitution}  and weakening, the sequent $\Gamma', x\ff{A} y, y:
A \vdash y: B, \Delta$ can be derived by  a derivation of the same
height as $u: A \vdash u: B$. Thus, the cut is replaced as follows
\[
\begin{prooftree}
\[
\[
\Gamma', x\ff{A} y, y:A \vdash y:B,\Delta \quad \Gamma', x\ff{A}
y, y: B \vdash \Delta \justifies \Gamma', x\ff{A} y, y: A \vdash
\Delta \using (cut)
\]
\justifies
 \Gamma', x\ff{A} y, x\ff{A} y \vdash \Delta
\using \rm (ID)
\]
\justifies \Gamma', x\ff{A} y, \vdash \Delta \using \rm (ContrL)
\end{prooftree}
\]

\item[(e)]
$F= x\ff{A} x$ is introduced on the left by (MP) rule, and it is
introduced on the right by (EQ). Thus we have
\[
\begin{prooftree}
\[\Gamma \vdash x: A,\Delta'
\justifies \Gamma\vdash \Delta', x\ff{A} x \using \rm (MP)
\]
\[u: A \vdash u: B \quad u: B \vdash u: A
\justifies \Gamma, x\ff{A} x\vdash \Delta', x \ff{B} x \using (EQ)
\]
\justifies  \Gamma\vdash \Delta',  x \ff{B} x \using \rm (cut)
\end{prooftree}
\]
where $\Delta = \Delta',   x \ff{B} x$. By Lemma \ref{lemma label
substitution} and weakening, the sequent $\Gamma, x: A \vdash x:
B, \Delta^{'}$ can be derived by a derivation of the same height
as $u: A \vdash u: B$. Thus the cut is replaced as follows:
\[
\begin{prooftree}
\[\Gamma \vdash x: A,\Delta'  \quad  \Gamma, x: A \vdash \Delta', x: B
\justifies \Gamma \vdash \Delta', x: B \using (cut)
\]
\justifies \Gamma \vdash \Delta', x\ff{B} x \using \rm (MP)
\end{prooftree}
\]

\item[(f)]
$F= x\ff{A} x$ is introduced on the right by (MP) rule and on the
left by (ID). Thus we have
\[
\begin{prooftree}
\[
\Gamma \vdash x: A,\Delta \justifies \Gamma \vdash \Delta, x\ff{A}
x \using (MP)
\]
\[
\Gamma, x:A \vdash \Delta \justifies \Gamma, x\ff{A} x \vdash
\Delta \using (ID)
\]
\justifies \Gamma\vdash \Delta \using \rm (cut)
\end{prooftree}
\]
We replace this cut by the following:
\[\irule{\Gamma \vdash x: A,\Delta \quad \Gamma, x:A \vdash \Delta}
{\Gamma\vdash \Delta}{(cut)}\]
\end{description}
\end{description}

\begin{teorema}[Completeness]\label{CComp}
If $A$ is valid in CK \{+MP\}\{+ID\}, then $\vdash x: A$ is
derivable in the respective SeqS system.
\end{teorema}

\noindent \emph{Proof}. We must show that the axioms are derivable
and that the set of derivable formulas is closed under (Modus
Ponens), (RCEA), and (RCK). A derivation of axioms (ID) and (MP)
is shown in examples \ref{ex1} and \ref{ex2} respectively.\\Let us
examine the other axioms.\\For (Modus Ponens), suppose that
$\vdash x: A \ri B$ and $\vdash x: A$ are derivable. We easily
have that $x: A \ri B, x: A\vdash x: B$ is derivable too. Since
cut is admissible, by two cuts we obtain $\vdash x: B$.\\For
(RCEA), we have to show that if $A \leftrightarrow B$ is
derivable, then also $(A\Ri C) \leftrightarrow (B\Ri C)$ is so.
The formula $ A \leftrightarrow B$ is an abbreviation for $(A \ri
B) \land (B \ri A)$. Suppose that $\vdash x: A \ri B$ and $\vdash
x: B \ri A$ are derivable, we can derive $x:B \Ri C \vdash x:A \Ri
C$ as follows: (the other half is symmetrical).


\[\begin{prooftree}
    \[
        \[
        x:A \vdash x:B \quad x:B \vdash x:A
     \justifies
        x \ff{B} y \vdash  x \ff{A} y,y:C
        \using (EQ)
        \]
        x \ff{B} y, y: C \vdash   y:C
     \justifies
        x \ff{B} y, x: A\Ri C \vdash y:C
        \using (\Ri L)
    \]
    \justifies
x: A\Ri C  \vdash  x: B\Ri C \using (\Ri R)
\end{prooftree}
\]


\noindent For (RCK), suppose that (1) $\vdash x:B_1 \wedge B_2
\dots \land B_n \rightarrow C$, it must be derivable also $x:
B_1,\ldots,x: B_n \vdash x: C$. We set $\Gamma_i = x:A\Ri
B_i,x:A\Ri B_{i+1},\dots x:A\Ri B_n$, for $1\leq i \leq n$. Then
we have (we  omit side formulas in $x \ff{A} y\vdash x \ff{A} y$):
\[
\begin{prooftree}
    \[  x \ff{A} y\vdash x \ff{A} y
        \[
            x \ff{A} y\vdash x \ff{A} y \quad x: B_1,\ldots,x: B_n \vdash x: C
            \justifies
        \shortstack{$x \ff{A} y, x: A\Ri B_n, x: B_1,\ldots,x: B_{n-1}\vdash y: C$\\
            $\vdots$\\
            $x \ff{A} y,\Gamma_2, y:B_1\vdash y:C$}
            \using (\Ri L)
        \]
    \justifies
        x \ff{A} y, x: A\Ri B_1, x: A\Ri B_2, \dots, x:A\Ri B_n \vdash y: C
    \using (\Ri L)
    \]
\justifies x: A\Ri B_1, x: A\Ri B_2, \dots, x:A\Ri B_n \vdash x: A
\Ri C \using (\Ri R)
\end{prooftree}
\]

\finedim

\section{Proof-theoretical Analysis of SeqS}
In this section we analyze the sequent calculus SeqS in order to
obtain a decision procedure for our conditional systems CK, CK+ID,
CK+MP and CK+MP+ID. In particular, we show that the contraction
rules (Contr L) and (Contr R) can be eliminated in SeqCK and
SeqID, but they cannot in SeqMP and SeqID+MP; however, in the last
two systems one can control the application of these rules, in
order to get a terminating calculus. Using the contraction rule in
a controlled way is essential to prove the existence of a decision
procedure for the associated logics; in fact, without this
control, one can apply contraction duplicating arbitrarily any
formula in the sequent.

First of all, we prove that we can eliminate the weakening rules.

\begin{teorema}[Elimination of weakening rules]\label{eliminazione weakening}
  Let $\Gamma$ $\prova$ $\Delta$ be a sequent  derivable in
  SeqS. Then $\Gamma$ $\prova$ $\Delta$  has a derivation in SeqS with no application
   of (Weak L) and   (Weak R).
\end{teorema}
  \emph{Proof}. By induction on the height of the proof tree.
\finedim

Now we introduce the notion of \emph{regular sequent}.
Intuitively, regular sequents are those sequents whose set of
transitions in the antecedent forms a \emph{forest}. As we show in
Theorem \ref{sequenti regolari} below, any sequent in a proof
beginning with a sequent of the form $\prova x_0: D$, for an
arbitrary formula $D$, is regular. For this reason, we will
restrict our concern to regular sequents.

We define the multigraph $\GI$ of the transition formulas in the
antecedent of a sequent:
\begin{definition}[Multigraph of
transitions $\GI$]
  Given a sequent $\Gamma  \prova \Delta$, where
  $\Gamma = \Gamma', T$ and $T$ is the
 multiset of transition formulas and $\Gamma'$ does   not contain
 transition formulas, we define  the multigraph $\GI = <V, E>$ associated
 to $\Gamma  \prova \Delta$ with vertexes $V$ and edges $E$.
  $V$ is the set of
 labels occurring in $\Gamma \prova \Delta$ and
   $<x, y> \appartiene E$ whenever $x \trans{F} y \in T$.
\end{definition}

\begin{definition}[Regular sequent]
  A sequent $\Gamma \prova \Delta$  is called
  regular   if its associated multigraph of transitions $\GI$ is a
  forest. In particular, there is at most one link between
  two vertexes and there are no loops.
\end{definition}

We can observe that we can always restrict our concern to regular
sequents, since we have the following theorem:

\begin{teorema}[Proofs with regular sequents]\label{sequenti regolari}
  Every proof tree beginning with a sequent $\prova x_0: D$ and obtained by
  applying SeqS's rules, contains only regular sequents.
\end{teorema}

\noindent \emph{Proof}. First, we show that $\GI$ is a
\emph{graph}, i.e. there is at most one link between two vertexes.
This can be seen by an easy inductive argument: $\prova x_0: D$
obviously respects this condition. Consider an arbitrary $\Gamma
\prova \Delta$ which respects this condition, ($\cond$ R) is the
only rule of the calculus which introduces, looking backward, a
transition formula in the antecedent of the sequent to which it is
applied. In particular, ($\cond$ R) with principal formula $x: A
\cond B$ introduces a transition $x \trans{A} y$ where $y$ is a
"new label", then there cannot be another transition $x \trans{F}
y$ in the antecedent.\\To see that $\GI$ is a forest, again we do
a simple inductive argument: the graph associated to $\prova x_0:
D$ is certainly a forest ($\GI$=$<$\{$x_0$\},$\vuoto>$, which is a
tree); consider a rule application which has $\Gamma_1 \prova
\Delta_1$ and $\Gamma_2 \prova \Delta_2$ as premises and $\Gamma
\prova \Delta$ as a conclusion, and assume by induction hypothesis
that the graph $\GI$ associated to $\Gamma \prova \Delta$ is a
forest. It is easy to observe that applying any rule of SeqS, the
graphs $\GI_1$ and $\GI_2$, associated to $\Gamma_1 \prova
\Delta_1$ and $\Gamma_2 \prova \Delta_2$ respectively, are
forests. (Contr L), (Contr R), (Weak L), (Weak R), ($\cond$ L),
($\imp$ L), ($\imp$ R) and (MP) do not modify the graph $\GI$;
($\cond$ R) adds a transition $x \trans{F} y$ in the initial
forest, but $y$ is a "new" label as discussed above, thus $\GI_1$
is still a forest obtained by adding a new vertex and a new edge;
consider the (ID) rule:
\[
  \begin{prooftree}
    \Gamma, y: A \prova \Delta
    \justifies \Gamma, x \trans{A} y \prova \Delta
    \using (ID)
  \end{prooftree}
\]
The application of (ID) deletes the edge $<x, y>$ from $\GI$, and
$\GI_1$ is still a forest. When (EQ) is applied, the calculus
tries to find two derivations starting with only one label $u$ and
no transitions; therefore, $\GI_1$ and $\GI_2$ are trees
$<$\{$u$\}, $\vuoto>$.
 \finedim

\noindent As mentioned above, we restrict from now on our
attention to regular sequents. In the following, we prove some
elementary properties of regular sequents.

\begin{teorema}[Property of ($\cond$ L)]\label{Proprieta di cond L}
  Let the sequent
    \begin{center}
      $\Gamma$ $\prova$ $\Delta$, $x \trans{A} y$
    \end{center}
  with $x \diverso y$, be derivable in SeqS, then one of the following sequents:
  \begin{enumerate}
    \item $\Gamma$ $\prova$ $\Delta$
    \item $x$ $\trans{F}$ $y$ $\prova$ $x$ $\trans{A}$ $y$, where $x$ $\trans{F}$
    $y$ $\appartiene$ $\Gamma$
  \end{enumerate}
  is also derivable in SeqS .
\end{teorema}
\emph{Proof}. (EQ) is the only SeqS's rule which operates on a
transition formula on the right hand side (consequent) of a
sequent. Thus, we have to consider only two cases, analyzing the
proof tree of $\Gamma$ $\prova$ $\Delta$, $x \trans{A} y$:
\begin{enumerate}
  \item $x$ $\trans{A}$ $y$ is introduced by weakening: in this
  case, $\Gamma$ $\prova$ $\Delta$ is derivable;
  \item $x$ $\trans{A}$ $y$ is introduced by the (EQ) rule: in this
  case, another transition $x$ $\trans{F}$ $y$ \emph{must} be in
  $\Gamma$, in order to apply (EQ). To see this, observe that
  the only rule that could introduce a transition formula (looking
  backward) in the antecedent of a sequent is ($\cond$ R), but it
  can only introduce a transition of the form $x$ $\trans{F}$ $z$, where
  \emph{z does not occur in that sequent} (it is a \emph{new}
  label), thus it cannot introduce the transition $x$ $\trans{F}$
  $y$.\\The (EQ) rule is only applied to transition formulas:
  \[
    \begin{prooftree}
      u: F \prova u: A \quad \quad \quad u: A \prova u: F
      \justifies x \trans{F} y \prova x \trans{A} y \using (EQ)
    \end{prooftree}
  \]
  therefore we can say that $x$ $\trans{F}$ $y$ $\prova$ $x$ $\trans{A}$
  $y$ is derivable in SeqS.
  \end{enumerate}
\finedim

Notice that this theorem holds for all the systems SeqS, but only
if $x \diverso y$. In SeqMP and SeqID+MP the (MP) rule operates on
transitions in the consequent, although on transitions like $x
\trans{A} x$. In this case the theorem does not hold, as shown by
the following counterexample:

\[
\begin{prooftree}
  x: A \prova x: A, x: B
  \justifies  x: A \prova x \trans{A} x, x: B
  \using (MP)
\end{prooftree}
\]

\noindent for $A$ and $B$ arbitrary. The sequent $x: A \prova x
\trans{A} x, x: B$ is derivable in SeqMP, but $x: A \prova x: B$ is not derivable in this system
and the second condition is not applicable (no
transition formula occurs in the antecedent).\\The first hypothesis of the theorem ($x \diverso y$)
excludes this situation.

\begin{teorema}[Elimination of the contraction rules on transition
formulas]
  Given a sequent $\Gamma$ $\prova$ $\Delta$, derivable in SeqS,
  there is a proof tree with no applications of (Contr L) and
  (Contr R) on transition formulas.
\end{teorema}
\emph{Proof} Consider one maximal\footnote{Of maximal distance
from the root of the proof tree.} contraction of the proof tree:
it can be eliminated as follows.\\In SeqCK a transition formula
$x$ $\trans{A}$ $y$ can be only introduced by the (EQ) rule
(looking forward):
  \[
    \begin{prooftree}
      \[
          u: B \prova u: A \quad \quad \quad \quad u: A \prova u:
          B
            \justifies \shortstack{
            $\Gamma^{''}$, \emph{x $\trans{B}$ y} $\prova$
            $\Delta^{''}$, \emph{x $\trans{A}$ y, x $\trans{A}$
            y}\\ $\Pi_1$ \\ $\Gamma^{'}$ $\prova$ $\Delta^{'}$, \emph{x $\trans{A}$ y, x $\trans{A}$
            y}
            }
             \using (EQ)
      \]
      \justifies \Gamma^{'} \prova \Delta^{'}, x \trans{A} y
      \using (Contr R)
    \end{prooftree}\\
  \]
  We can obtain the following proof erasing the contraction step:\\
  \[
    \begin{prooftree}
          u: B \prova u: A \quad \quad \quad \quad u: A \prova u:
          B
      \justifies
        \shortstack{
          $\Gamma^{''}$, \emph{x $\trans{B}$ y} $\prova$ $\Delta^{''}$, \emph{x $\trans{A}$ y}\\
          $\Pi_1^{'}$\\
          $\Gamma^{'}$ $\prova$ $\Delta^{'}$, \emph{x $\trans{A}$ y}
      }
      \using (EQ)
    \end{prooftree}\\
  \]
where $\Pi_1^{'}$ is obtained by removing an occurrence of \emph{x
$\trans{A}$ y} on the right side of every sequent of $\Pi_1$.\\In
SeqID and SeqID+MP we can have proofs like the following one:
  \[
    \begin{prooftree}
      \[
      \[
        \shortstack{
          $\Pi_3$\\
          $\Gamma^{'''}$, \emph{y: A} $\prova$
          $\Delta^{'''}$
        }
        \justifies
          \shortstack{
          $\Gamma^{'''}$, \emph{x $\trans{A}$ y} $\prova$
          $\Delta^{'''}$\\
          $\Pi_2$\\
          $\Gamma^{''}$, \emph{x $\trans{A}$ y, y: A} $\prova$
          $\Delta^{''}$
          }
        \using (ID)
      \]
        \justifies \shortstack{
          $\Gamma^{''}$, \emph{x $\trans{A}$ y, x $\trans{A}$ y} $\prova$
          $\Delta^{''}$\\
          $\Pi_1$\\
          $\Gamma^{'}$, \emph{x $\trans{A}$ y, x $\trans{A}$ y} $\prova$
          $\Delta^{'}$
        } \using (ID)
      \]
      \justifies \Gamma^{'}, x \trans{A} y \prova \Delta^{'}
      \using (Contr L)
    \end{prooftree}\\
  \]
By the permutability of (ID) over the other rules of SeqS, we can
have:
  \[
    \begin{prooftree}
      \[
        \[
            \shortstack{
            $\Pi_3$\\
            $\Gamma^{'''}$,\emph{y: A} $\prova$
            $\Delta^{'''}$\\
            $\Pi_1^{'}$ \\
            $\Gamma^{'}$,\emph{y: A, y: A} $\prova$
            $\Delta^{'}$
            }
          \justifies \Gamma^{'}, y: A, x \trans{A} y \prova
          \Delta^{'}\using (ID)
        \]
        \justifies \Gamma^{'}, x \trans{A} y, x \trans{A} y \prova \Delta^{'}
        \using (ID)
      \]
      \justifies \Gamma^{'}, x \trans{A} y \prova \Delta^{'}
      \using (Contr L)
   \end{prooftree}\\
  \]
where $\Pi_1^{'}$ is obtained from $\Pi_1$ and $\Pi_2$ by
permuting the two applications of (ID). We can then eliminate the
contraction on the transition formula, introducing an application
of (Contr L) on the subformula \emph{y: A}:
  \[
    \begin{prooftree}
      \[
            \shortstack{
            $\Pi_3$\\
            $\Gamma^{'''}$,\emph{y: A} $\prova$
            $\Delta^{'''}$\\
            $\Pi_1^{'}$ \\
            $\Gamma^{'}$,\emph{y: A, y: A} $\prova$
            $\Delta^{'}$
            }
          \justifies \Gamma^{'}, y: A \prova
          \Delta^{'}\using (Contr L)
        \]
        \justifies \Gamma^{'}, x \trans{A} y \prova \Delta^{'}
        \using (ID)
   \end{prooftree}
  \]
In SeqMP and SeqID+MP we can eliminate a contraction step on a
transition formula $x \trans{A} x$ in the consequent of a sequent
by replacing it with an application of (Contr R) on the subformula
\emph{x: A} in a similar way to the case of (ID).\\In all the SeqS
calculi, if a transition formula is introduced by (implicit)
weakening, the contraction is eliminated by eliminating that
weakening.

\finedim

\subsection{Elimination of (Contr R) on Conditional Formulas}
\begin{figure}
{\centerline{\includegraphics[angle=0,width=5.0in]{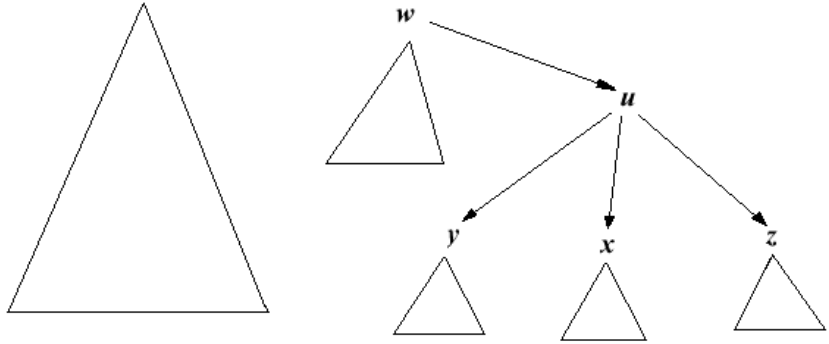}}}
\caption{The forest $\GI$ of a sequent; $w$ is predecessor of $x$,
$x$ is successor of $w$, $u$ is father of $x$ and $x$ is son of
$u$.}\label{Foresta delle transizioni}
\end{figure}

In this subsection we show that SeqS calculi are still complete
without the (Contr R) rule applied to conditional formulas
\emph{x: A $\cond$ B}. The elimination of the right contraction on
conditionals is a direct consequence of the so-called
\emph{disjunction property} for conditional formulas: if ($A_1
\cond B_1$) $\orr$ ($A_2 \cond B_2$) is valid, then either ($A_1
\cond B_1$) or ($A_2 \cond B_2$) is valid too. This property
follows an important proposition, which does not hold for all
sequents, but only for non-$x$-branching sequents, i.e. those
sequents which do not create a branching in $x$ or in a
predecessor of $x$. Let us introduce some essential definitions.
\begin{definition}[Predecessor and successor, father and son]
  Given a sequent $\Gamma^{'}, T \prova \Delta$, where all the transitions in the antecedent are in $T$, we say
  that a world $w$ is a predecessor of a world $x$ if there is a path from $w$ to $x$ in the graph
  of transitions $\GI=<V, E>$ of the sequent. In this case, we also say
  that $x$ is a successor of $w$. If $<w, x> \appartiene E$, we
  say that $w$ is the father of $x$ and that $x$ is a son of $w$.
\end{definition}
As we mentioned above, the graph of transitions forms a forest, as
shown in Figure \ref{Foresta delle transizioni}.

\begin{definition}[Positive and negative occurrences of a formula]
  Given a formula $A$, we say that:
  \begin{list}{-}{}
    \item $A$ occurs positively in $A$;
    \item if a formula $B$ $\imp$ $C$ occurs positively
    (negatively) in $A$, then $C$ occurs positively (negatively)
    in $A$ and $B$ occurs negatively (positively) in $A$;
    \item if a formula $B$ $\cond$ $C$ occurs positively
    (negatively) in $A$, then $C$ occurs positively (negatively) in
    $A$.
  \end{list}
  A formula $F$ occurs positively (negatively) in a multiset
  $\Gamma$ if $F$ occurs positively (negatively) in some formula
  $G$ $\appartiene$ $\Gamma$.
\end{definition}
 Now we introduce the
definition of $x$-branching formula. Intuitively, $\B$($x, T$)
contains formulas that create a branching in $x$ or in a
predecessor of $x$ according to $T$. $\B$($x, T$) also contains
the conditionals $u: A \cond B$ such that $T \prova u \trans{A} v$ and $B$ creates a branching
in $x$ (i.e. $v=x$) or in a predecessor $v$ of $x$.
\begin{definition}[$x$-branching formulas]
  Given a multiset of transition formulas $T$, we define the set
  of $x$-branching formulas, denoted with $\BB$($x$, $T$), as
  follows:
  \begin{list}{-}{}
    \item $x: A \imp B \appartiene \B$($x, T$);
    \item $u: A \imp B \appartiene \B$($x, T$) if
    $T \prova u \trans{F} x$ for some formula
    $F$;
    \item $u: A \cond B \appartiene \B$($x, T$) if
    $T \prova u \trans{A} v$ and $v: B \appartiene
    \B$($x, T$).
  \end{list}
\end{definition}

We also introduce the notion of $x$-branching sequent.
Intuitively, we say that $\Gamma \prova \Delta$ is $x$-branching
if it contains an $x$-branching formula occurring positively in
$\Gamma$ or if it contains an $x$-branching formula occurring
negatively in $\Delta$. Since in systems containing (ID) a
transition $u \trans{F} v$ in the antecedent can be derived from
$v: F$ and $v:
  F$ can be $x$-branching, we impose that a sequent $\Gamma^{'}, u \trans{F} v \prova \Delta$ is
  $x$-branching if $\Gamma^{'}, v: F \prova \Delta$ is
  $x$-branching; for the same reason, in systems containing (MP) we impose that a sequent
  $\Gamma \prova \Delta^{'}, u \trans{F} u$ is
  $x$-branching if $\Gamma \prova \Delta^{'}, u: F$ is
  $x$-branching.\\In systems containing (MP) we also impose that a
  sequent $\Gamma^{'}, w: A \cond B \prova \Delta$ is
  $x$-branching if the
  sequent $\Gamma^{'} \prova \Delta, w \trans{A} w$ is derivable and $w$ is a predecessor of $x$ (or
  $w=x$),
  since $w: A$ can introduce $x$-branching formula(s) in the
  sequent.

\begin{definition}[$x$-branching sequents]
Given a sequent $\Gamma$ $\prova$ $\Delta$, we denote by
$\Gamma^{'}$ the labelled formulas in $\Gamma$ and by $T$ the
transition formulas in $\Gamma$, so that $\Gamma$=$\Gamma^{'}$,
$T$. To define when a sequent $\Gamma \prova \Delta$ is
$x$-branching according to each system, we consider the following
conditions:
\begin{enumerate}
  \item a formula $u: F \appartiene \B$($x, T$) occurs positively in $\Gamma$;
  \item a formula $u: F \appartiene \B$($x, T$) occurs negatively in $\Delta$.
  \item $T$=$T^{'}, u \trans{F} v$ and the sequent $\Gamma^{'}, T^{'}, v: F \prova \Delta$ is
  $x$-branching;
  \item $u \trans{F} u \appartiene
  \Delta$ and the sequent $\Gamma \prova \Delta^{'}, u: F$ is
  $x$-branching ($\Delta=\Delta^{'},u \trans{F} u$);
  \item a formula $w: A \cond B \appartiene \Gamma$, $w$ is a predecessor of
  $x$ in the graph $\GI$ of transitions or $w=x$ and $\Gamma^{''} \prova \Delta, w \trans{A} w$ is derivable,
  where $\Gamma=\Gamma^{''}, w: A \cond B$.
\end{enumerate}

\noindent We say that $\Gamma \prova \Delta$ is $x$-branching for
each system if the following combinations of the previous
conditions hold:
\begin{list}{-}{}
  \item CK: 1, 2
  \item CK+ID: 1, 2, 3
  \item CK+MP: 1, 2, 4, 5
  \item CK+MP+ID: 1, 2, 3, 4, 5
\end{list}
\label{sequente x-branching}
\end{definition}

\begin{figure}
{\centerline{\includegraphics[angle=0,width=5.0in]{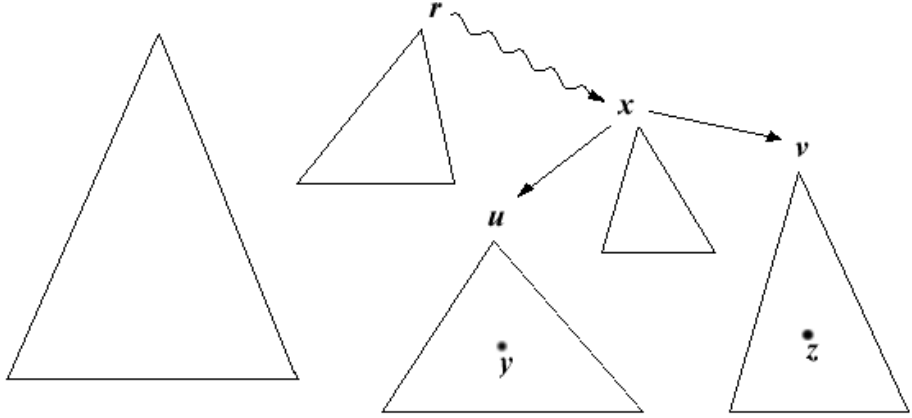}}}
 \caption{The forest
$\GI$ of transitions used to prove the disjunction property.}
\label{Foresta per provare la proposizione}
\end{figure}

The disjunction property characterizes \emph{only} the sequents
that are not \emph{x}-branching. To prove the disjunction
property, we need to consider a more general setting; namely, we
shall consider a sequent of the form
  $\Gamma \prova \Delta, y: A, z: B$, whose
   forest of transitions   has the form represented in Figure
  \ref{Foresta per provare la proposizione}, i.e. it has one subtree with root $u$ and another
  subtree with root $v$, with $u \diverso v$;
$y$ is a member of the tree with root $u$ and $z$ is a member of
the tree with root $v$; $x$ is the father of $u$ and $v$ and the tree
containing $x$ has root $r$.

We need some more definitions. In particular, given a sequent
$\Gamma \prova \Delta$ with its associated forest $\GI=<V, E>$,
consider a label $k$ contained in the tree with root $r$. We
define the set $T_k^{\pr}$ of the labels contained in the tree of
$\GI$ with root $k$ and the sets $\Gamma_k^{\pr}$ and
$\Delta_k^{\pr}$, containing all the formulas of $\Gamma$ and
$\Delta$ whose labels are in the tree of $\GI$ with root $k$. We
also define the set $T^{*}_k$ of the labels contained in the tree
of $\GI$ with root $k$ or in a path from $r$ to $k$, and the sets
$\Gamma_k^{*}$ and $\Delta_k^{*}$, containing all the formulas of
$\Gamma$ and $\Delta$ whose labels are in the tree of $\GI$ with
root $k$ or on the path from $r$ to $k$.

\begin{definition}[$T_k^{\pr}$]
  $T_k^{\pr}$ is the set of labels in the tree of $\GI$ with root $k$; more precisely:
\begin{list}{-}{}
  \item $k$ $\appartiene$ $T^{\pr}_k$
  \item if $<u, w>$ $\appartiene$
  $E$ and $u$ $\appartiene$ $T^{\pr}_k$, then $w$ $\appartiene$
  $T^{\pr}_k$.
\end{list}
\end{definition}

\begin{definition}[$T_k^{*}$]
  $T_k^{*}$ is the set of labels in the tree of $\GI$ with root $k$ or on a path from $r$\footnote{The label $r$
  is the root of the tree containing $k$. It could be $r$=$k$.} to $k$; more precisely:
  \begin{center}
    $T_k^{*}=T^{\pr}_k \unione P_k$
  \end{center}
  where $P_k$ it the set of labels on a path from $r$ to $k$:
\begin{list}{-}{}
  \item $k$ $\appartiene$ $P_k$
  \item if $<u, w>$ $\appartiene$
  $E$ and $w$ $\appartiene$ $P_k$, then $u$ $\appartiene$
  $P_k$.
\end{list}
\end{definition}

\begin{definition}[$\Gamma^{\pr}_k$]
  $\Gamma^{\pr}_k$ is the multiset of formulas of $\Gamma$ contained in the tree of $\GI$ with root
  $k$; if $w \trans{F} k \appartiene \Gamma$, then $w \trans{F} k \appartiene \Gamma^{\pr}_k$;
  more precisely:
  \begin{center}
      $\Gamma^{\pr}_k=$\{$u: F \appartiene \Gamma \tc u
      \appartiene T^{\pr}_k$\} $\unione$ \{$w \trans{F} u \appartiene \Gamma \tc w \appartiene
      T^{\pr}_k$ or $u \appartiene T^{\pr}_k$\}
  \end{center}
\end{definition}
\begin{definition}[$\Delta^{\pr}_k$]
  $\Delta^{\pr}_k$ is the multiset of formulas of $\Delta$ contained in the tree of $\GI$ with root
  $k$; if $w \trans{F} k \appartiene \Delta$, then $w \trans{F} k \appartiene \Delta^{\pr}_k$;
  more precisely:
  \begin{center}
      $\Delta^{\pr}_k=$\{$u: F \appartiene \Delta \tc u
      \appartiene T^{\pr}_k$\} $\unione$ \{$w \trans{F} u \appartiene \Delta \tc w \appartiene
      T^{\pr}_k$ or $u \appartiene T^{\pr}_k$\}
  \end{center}
\end{definition}

\begin{definition}[$\Gamma^{*}_k$]
  $\Gamma^{*}_k$ is the multiset of formulas of $\Gamma$ contained in the tree of $\GI$ with root
  $k$ or on a path from $r$ to $k$;
  more precisely:
  \begin{center}
      $\Gamma^{*}_k=\{w: F \appartiene \Gamma \tc w \appartiene
      T^{*}_k\} \unione \{w \trans{F} w^{'} \appartiene \Gamma \tc w^{'} \appartiene T^{*}_k\}$
  \end{center}
\end{definition}
\begin{definition}[$\Delta^{*}_k$]
  $\Delta^{*}_k$ is the multiset of formulas of $\Delta$ contained in the tree of $\GI$ with root
  $k$ or on a path from $r$ to $k$;
  more precisely:
  \begin{center}
      $\Delta^{*}_k=\{w: F \appartiene \Delta \tc w \appartiene
      T^{*}_k\} \unione \{w \trans{F} w^{'} \appartiene \Delta \tc w^{'} \appartiene T^{*}_k\}$
  \end{center}
\end{definition}
Now we have all the elements to prove the following:

\begin{proposizione}\label{proposizione 1 della tesi}
Given a sequent $\Gamma \prova \Delta,
y: A, z: B$ and its forest of transitions $\GI$, if it is
derivable in SeqS and has the following features:
    \begin{enumerate}
      \item $\GI$ is a forest of the form as shown in Figure \ref{Foresta per provare la proposizione} (thus $y$ is a
      member of the tree with root $u$  and $z$ is a
      member of the tree with root $v$, with $u \diverso v$; $u$ and $v$ are sons of $x$);
      \item $\Gamma \prova \Delta, y: A, z: B$ is not
      $x$-branching
    \end{enumerate}
    then one of the following sequents is derivable in SeqS:
    \begin{enumerate}
      \item $\Gamma^{*}_u$ $\prova$ $\Delta^{*}_u, y: A$
      \item $\Gamma^{*}_v$ $\prova$ $\Delta^{*}_v, z: B$
      \item $\Gamma$ - $(\Gamma^{\pr}_u$ $\unione$ $\Gamma^{\pr}_v)$ $\prova$
        $\Delta$ - $(\Delta^{\pr}_u$ $\unione$ $\Delta^{\pr}_v)$
    \end{enumerate}
\end{proposizione}

\noindent Moreover, the proofs of 1, 2 and 3 do not add any
contraction to the proof of $\Gamma \prova \Delta, y: A, z:
B$.\\\\ \emph{Proof}. By induction on the height of the proof tree
of the
 sequent $\Gamma \prova \Delta, y: A, z: B$. We present two
 examples, the other cases are left to the reader.
 \begin{enumerate}
   \item Consider the case where $y: A$ is a conditional formula $y: C \cond D$
   and is the principal formula of an application of the ($\cond$ R)
   rule. The proof tree of the sequent is ended by:
    \[
           \begin{prooftree}
             \Gamma, y \trans{C} k \prova \Delta, k: D, z: B
             \justifies \Gamma \prova \Delta, y: C \cond D, z: B
             \using (\cond R)
           \end{prooftree}\\
         \]
   We can apply the inductive hypothesis on the only premise of
   the ($\cond$ R) rule; in fact, $\Gamma, y \trans{C} k \prova \Delta, k: D, z:
   B$ is not $x$-branching. It could become $x$-branching as an
   effect of the introduction of $k: D$ and $y
   \trans{C} k$, but this is impossible since $k$ is a "new" label, then it is in the same
   tree of $y$ and not on a path to $x$. Applying the inductive hypothesis, we must consider the
   three possible situations:
   \begin{enumerate}
           \item (\emph{$\Gamma$, y $\trans{C}$ k}$)^{*}_u$ $\prova$
           $\Delta^{*}_u$, \emph{k: D} is derivable: it is easy to see that \emph{y $\trans{C}$ k}
           $\appartiene$ (\emph{$\Gamma$, y $\trans{C}$ k}$)^{*}_u$,
           since $u$ is a predecessor of $y$ and thus of $k$; then
           we obtain the following derivation:\\
             \[
               \begin{prooftree}
                 \Gamma^{*}_u, y \trans{C} k \prova \Delta^{*}_u, k: D
                 \justifies \Gamma^{*}_u \prova \Delta^{*}_u, y: C
                 \cond D \using (\cond R)
               \end{prooftree}\\
             \]

           \item (\emph{$\Gamma$, y $\trans{C}$ k}$)^{*}_v$ $\prova$
           $\Delta^{*}_v$, \emph{z: B} is derivable: \emph{k} is in the tree with root \emph{u}, thus \emph{y $\trans{C}$ k}
           $\not\appartiene$ (\emph{$\Gamma$, y $\trans{C}$
           k}$)^{*}_v$: we obtain that\\
           \[
             \Gamma^{*}_v \prova \Delta^{*}_v, z: B
           \]
           is derivable;

           \item (\emph{$\Gamma$, y $\trans{C}$ k}) - ((\emph{$\Gamma$, y $\trans{C}$ k}$)^{\pr}_u$ $\unione$
           (\emph{$\Gamma$, y $\trans{C}$ k}$)^{\pr}_v$) $\prova$
           $\Delta$ - ($\Delta^{\pr}_u$ $\unione$
           $\Delta^{\pr}_v$) is derivable: \emph{k} is in the subtree with root \emph{u}, thus \emph{y $\trans{C}$
           k} $\appartiene$ (\emph{$\Gamma$, y $\trans{C}$
           k}$)^{\pr}_u$ and then \emph{y $\trans{C}$
           k} $\not\appartiene$ (\emph{$\Gamma$, y $\trans{C}$ k}) -
           ((\emph{$\Gamma$, y $\trans{C}$ k}$)^{\pr}_u$ $\unione$
           (\emph{$\Gamma$, y $\trans{C}$ k}$)^{\pr}_v$), from which we obtain that\\
           \[
             \Gamma - (\Gamma^{\pr}_u \unione \Gamma^{\pr}_v)
             \prova \Delta - (\Delta^{\pr}_u \unione \Delta^{\pr}_v)
           \]
           is derivable.
   \end{enumerate}
   \item Let us now analyze the case where the principal
   formula of the sequent is a formula $w: F \appartiene \Delta$;
   the (Contr R) rule is applied to that formula, as shown below:
            \[
              \begin{prooftree}
                \Gamma \prova \Delta^{'}, w: F, w: F, y: A, z: B
                \justifies \Gamma \prova \Delta^{'}, w: F, y: A, z: B \using (Contr
                R)
              \end{prooftree}\\
            \]
    We can obviously apply the inductive hypothesis on the
    premise, obtaining the three following alternatives:
            \begin{enumerate}
              \item \emph{$\Gamma^{*}_u$ $\prova$}
              (\emph{$\Delta^{'}$, w: F, w: F}$)^{*}_u$, \emph{y:
              A} is derivable: if \emph{w} $\appartiene$ $T^{*}_u$,
              then the sequent $\Gamma^{*}_u$ $\prova$
              $\Delta^{'*}_u$, \emph{w: F, w: F, y: A} is derivable, as \emph{w: F} occurs
              in ($\Delta^{'}$, \emph{w: F, w: F}$)^{*}_u$. Therefore we can obtain the following proof:\\
              \[
                \begin{prooftree}
                  \Gamma^{*}_u \prova \Delta^{'*}_u, w: F, w: F, y:
                  A \justifies \Gamma^{*}_u \prova \Delta^{'*}_u, w: F, y: A
                  \using (Contr R)
                \end{prooftree}
              \]
              and that's it, since
              $\Delta$=$\Delta^{'}$, \emph{w: F} and then $\Delta^{*}_u$=$\Delta^{'*}_u$, \emph{w:
              F}.\\If \emph{w} $\not\appartiene$ $T^{*}_u$, \emph{w: F} is not member of the multiset
              ($\Delta^{'}$, \emph{w: F, w:
              F}$)^{*}_u$,
              then the sequent $\Gamma^{*}_u$ $\prova$
              $\Delta^{'*}_u$, \emph{y: A} is derivable by the inductive hypothesis,
              from what we can conclude since $\Delta$=$\Delta^{'}$,
              \emph{w: F} and then $\Delta^{*}_u$=$\Delta^{'*}_u$.

              \item $\Gamma^{*}_v$ $\prova$
              ($\Delta^{'}, w: F, w: F$ $)^{*}_v$, $z:
              B$ is derivable: the proof is similar to the
              previous one and $\Gamma^{*}_v \prova$ ($\Delta^{'}, w: F$ $)^{*}_v$, $z: B$ is derivable .

              \item \emph{$\Gamma$ -} (\emph{$\Gamma^{\pr}_u$ $\unione$
              $\Gamma^{\pr}_v$}) $\prova$ (\emph{$\Delta^{'}$, w: F, w:
              F}) - ((\emph{$\Delta^{'}$, w: F, w: F}$)^{\pr}_u$ $\unione$
              (\emph{$\Delta^{'}$, w: F, w: F}$)^{\pr}_v$) is derivable: if \emph{w} $\appartiene$ $T^{\pr}_u$
              $\unione$ $T^{\pr}_v$, then $\Gamma$ - ($\Gamma^{\pr}_u$ $\unione$
              $\Gamma^{\pr}_v$) $\prova$ $\Delta^{'}$ - ($\Delta^{'\pr}_u$ $\unione$
              $\Delta^{'\pr}_v$) is derivable, and we can conclude the proof as $\Delta$=$\Delta^{'}$, \emph{w: F}, but \emph{w:
              F} $\appartiene$ $\Delta^{\pr}_u$ $\unione$
              $\Delta^{\pr}_v$, then it is not member of the difference, thus $\Delta$ - ($\Delta^{\pr}_u$ $\unione$
              $\Delta^{\pr}_v$)=$\Delta^{'}$ - ($\Delta^{'\pr}_u$ $\unione$
              $\Delta^{'\pr}_v$).\\If \emph{w: F} $\not\appartiene$
              $T^{\pr}_u$ $\unione$ $T^{\pr}_v$, then $\Gamma$ - ($\Gamma^{\pr}_u$ $\unione$
              $\Gamma^{\pr}_v$) $\prova$ ($\Delta^{'}$, \emph{w: F, w: F}) - ($\Delta^{'\pr}_u$ $\unione$
              $\Delta^{'\pr}_v$) is derivable by the inductive hypothesis, from what we can have:\\
              \[
                \begin{prooftree}
                  \Gamma - (\Gamma^{\pr}_u \unione \Gamma^{\pr}_v)
                  \prova \Delta^{'} - (\Delta^{'\pr}_u \unione
                  \Delta^{'\pr}_v), w: F, w: F
                  \justifies \Gamma - (\Gamma^{\pr}_u \unione \Gamma^{\pr}_v)
                  \prova \Delta^{'} - (\Delta^{'\pr}_u \unione
                  \Delta^{'\pr}_v), w: F \using (Contr R)
                \end{prooftree}
              \]
              and we can conclude the proof, since $\Delta$=$\Delta^{'}$,
              \emph{w: F} and \emph{w} $\not\appartiene$ $T^{\pr}_u$ $\unione$
              $T^{\pr}_v$; then, we observe that $\Delta$ - ($\Delta^{\pr}_u$ $\unione$
              $\Delta^{\pr}_v$)=$\Delta^{'}$ - ($\Delta^{'\pr}_u$ $\unione$
              $\Delta^{'\pr}_v$), \emph{w: F}.

              In this case, we introduce a (Contr R) to prove
              the Proposition; however, this contraction is already in the
              proof tree of the initial sequent, thus we do not \emph{add} any
              contraction on it (we use the same contraction on $w: F$).

            \end{enumerate}

 \end{enumerate}

\finedim

\begin{teorema}[Disjunction
property]
  Given a non $x$-branching sequent\\
\[
  \Gamma \prova \Delta, x: A_1 \cond B_1, x: A_2 \cond B_2
\]
derivable in SeqS with a derivation $\Pi$, one of the following
sequents:
\begin{enumerate}
  \item $\Gamma$ $\prova$ $\Delta$, $x: A_1 \cond B_1$
  \item $\Gamma$ $\prova$ $\Delta$, $x: A_2 \cond B_2$
\end{enumerate}
is derivable in SeqS by a proof tree which does not add any
application of the contraction rules (Contr L) and (Contr R) to
$\Pi$.
\end{teorema}
\emph{Proof}. The sequent $\Gamma \prova \Delta, x: A_1 \cond B_1,
x: A_2 \cond B_2$ is derivable in SeqS, then we can find a
derivation $\Pi$ of it; the two conditional formulas can be
introduced (looking forward) in two ways:
  \begin{enumerate}
    \item by weakening;
    \item by the application of the ($\cond$ R) rule.
  \end{enumerate}
In case 1 suppose that \emph{x: $A_1$ $\cond$ $B_1$} is introduced
by weakening (by Theorem \ref{eliminazione weakening} we can only
consider implicit weakenings): the proof is ended by erasing all
the instances of \emph{x: $A_1$ $\cond$ $B_1$} introduced by
weakening in $\Pi$, obtaining a proof of $\Gamma$ $\prova$
$\Delta$, \emph{x: $A_2$ $\cond$ $B_2$}.\\In case 2 both the
conditional formulas are introduced by an application of the
($\cond$ R) rule; by the permutability of this rule over all the
others, we can consider a proof tree ending as follows:
\[
  \begin{prooftree}
    \[
      \Gamma, x \trans{A_1} y, x \trans{A_2} z \prova
        \Delta, y: B_1, z: B_2
      \justifies \Gamma, x \trans{A_1} y \prova \Delta, y: B_1, x: A_2
         \cond B_2 \using (\cond R)
    \]
    \justifies \Gamma \prova \Delta, x: A_1 \cond B_1, x: A_2
    \cond B_2 \using (\cond R)
  \end{prooftree}
\]\\
The sequent $\Gamma, x \trans{A_1} y, x \trans{A_2} z \prova
\Delta, y: B_1, z: B_2$ respects all the conditions to apply the
Proposition \ref{proposizione 1 della tesi} ($y$ and $z$ are "new"
labels), then we have that one of the following sequents:
\begin{enumerate}
  \item ($\Gamma$, \emph{x $\trans{A_1}$ y, x $\trans{A_2}$
  z}$)^{*}_y$ $\prova$ $\Delta^{*}_y$, \emph{y: $B_1$}
  \item ($\Gamma$, \emph{x $\trans{A_1}$ y, x $\trans{A_2}$
  z}$)^{*}_z$ $\prova$ $\Delta^{*}_z$, \emph{z: $B_2$}
  \item ($\Gamma$, \emph{x $\trans{A_1}$ y, x $\trans{A_2}$
  z}) - (($\Gamma$, \emph{x $\trans{A_1}$ y, x $\trans{A_2}$
  z}$)^{\pr}_y$ $\unione$ ($\Gamma$, \emph{x $\trans{A_1}$ y, x $\trans{A_2}$
  z}$)^{\pr}_z$) $\prova$ $\Delta$ - ($\Delta^{\pr}_y$ $\unione$ $\Delta^{\pr}_z$)
\end{enumerate}
is derivable in SeqS without adding any application of the
contraction rules.\\In all these cases we can proof the
disjunction property:
\begin{enumerate}
  \item \emph{z} is not in the tree with root \emph{y},
  and is not on a path towards \emph{y}, then the transition formula \emph{x $\trans{A_2}$ z}
  is not member of the multiset ($\Gamma$, \emph{x $\trans{A_1}$ y, x $\trans{A_2}$ z}$)^{*}_y$; the sequent
   $\Gamma^{*}_y$, \emph{x $\trans{A_1}$ y} $\prova$ $\Delta^{*}_y$, \emph{y: $B_1$} is then derivable, from
   what we have a proof:\\
  \[
    \begin{prooftree}
    \[
    \Gamma^{*}_y, x \trans{A_1} y \prova \Delta^{*}_y, y: B_1
    \justifies \Gamma^{*}_y \prova \Delta^{*}_y, x: A_1 \cond B_1
    \using (\cond R)
    \]
    \justifies \Gamma \prova \Delta, x: A_1 \cond B_1 \using
    (Weak)
    \end{prooftree}
  \]\\
  since $\Gamma^{*}_y$ $\incluso$ $\Gamma$ and $\Delta^{*}_y$
  $\incluso$ $\Delta$;

  \item symmetric to the previous case;

  \item we can observe that \emph{x $\trans{A_1}$ y} $\appartiene$ ($\Gamma$, \emph{x $\trans{A_1}$ y, x $\trans{A_2}$
  z}$)^{\pr}_y$, and that \emph{x $\trans{A_2}$ z} $\appartiene$ ($\Gamma$, \emph{x $\trans{A_1}$ y, x $\trans{A_2}$
  z}$)^{\pr}_z$; both the transition formulas are members of ($\Gamma$, \emph{x $\trans{A_1}$ y, x $\trans{A_2}$
  z}$)^{\pr}_y$ $\unione$ ($\Gamma$, \emph{x $\trans{A_1}$ y, x $\trans{A_2}$
  z}$)^{\pr}_z$ and then they are \emph{not} members of ($\Gamma$, \emph{x $\trans{A_1}$ y, x $\trans{A_2}$
  z}) - (($\Gamma$, \emph{x $\trans{A_1}$ y, x $\trans{A_2}$
  z}$)^{\pr}_y$ $\unione$ ($\Gamma$, \emph{x $\trans{A_1}$ y, x $\trans{A_2}$
  z}$)^{\pr}_z$). Therefore, the sequent
   $\Gamma$ - ($\Gamma^{\pr}_y$ $\unione$
  $\Gamma^{\pr}_z$) $\prova$ $\Delta$ - ($\Delta^{\pr}_y$ $\unione$
  $\Delta^{\pr}_z$) is derivable and, observing that $\Gamma$ - ($\Gamma^{\pr}_y$ $\unione$
  $\Gamma^{\pr}_z$) $\incluso$ $\Gamma$ and that $\Delta$ - ($\Delta^{\pr}_y$ $\unione$
  $\Delta^{\pr}_z$) $\incluso$ $\Delta$, we have the proof:\\
  \[
    \begin{prooftree}
    \[
      \Gamma - (\Gamma^{\pr}_y \unione \Gamma^{\pr}_z) \prova
      \Delta - (\Delta^{\pr}_y \unione \Delta^{\pr}_z)
      \justifies \Gamma \prova \Delta \using (Weak)
    \]
    \justifies \Gamma \prova \Delta, x: A_1 \cond B_1 \using (Weak
    R)
    \end{prooftree}
  \]
or\\
  \[
    \begin{prooftree}
    \[
      \Gamma - (\Gamma^{\pr}_y \unione \Gamma^{\pr}_z) \prova
      \Delta - (\Delta^{\pr}_y \unione \Delta^{\pr}_z)
      \justifies \Gamma \prova \Delta \using (Weak)
    \]
    \justifies \Gamma \prova \Delta, x: A_2 \cond B_2 \using (Weak
    R)
    \end{prooftree}
  \]
\end{enumerate}
\finedim

By the correctness and completeness of SeqS, it is easy to prove
the following corollary of the disjunction property:
\begin{corollario}
  If $(A \cond B)$ $\orr$ $(C \cond D)$ is valid in CK\{+MP\}\{+ID\},
  then either $A \cond B$ or $C \cond D$ is valid in CK\{+MP\}\{+ID\}.
\end{corollario}

Finally, we can prove the eliminability of the (Contr R) in SeqS
systems on conditional formulas as another corollary of the
disjunction property.
\begin{corollario}[Elimination of the (Contr R) rule on conditional formulas]
If $\vdash x_0: D$ is derivable in SeqS, then it has a proof where
there are no right contractions on conditional formulas.
\end{corollario}
\emph{Proof}. By permutation properties, a proof $\Pi$ ending with
\[\Gamma \vdash \Delta, x: A \Ri B, x: A \Ri B\]
can be transformed into a proof $\Pi'$, where all the rules
introducing $x$-branching formulas are permuted over the other
rules (i.e. they are applied at the bottom of the tree). As an
example, let the end sequent of $\Pi$ have the form
\begin{center}
$\Gamma, x: C \ri D\vdash \Delta, x: A\Ri B, x: A\Ri B$
\end{center}
We have that the lower sequent is $x$-branching, (at least) since
$x: C\ri D$.  We can permute  $\Pi$ so that the last step is the
introduction of the $x$-branching formula $x: C \ri D$ from the
two sequents:
\[\Gamma\vdash \Delta, x:C,  x: A\Ri B, x: A\Ri B~\mbox{and} ~\Gamma,x: D \vdash \Delta,  x: A\Ri B, x: A\Ri B.\]
We have decomposed the $x$-branching formula, if the two sequents
are  still $x$-branching we perform a similar permutation upwards,
so that at the end every branch of $\Pi'$ will contain a sequent
$\Gamma_i \vdash \Delta_i, x: A \Ri B, x: A \Ri B$, such that
$\Gamma_i, \Delta_i$ are no longer $x$-branching. Notice that if a
sequent $\Gamma, w: C \cond D\vdash \Delta, x: A\Ri B, x: A\Ri B$
is $x$-branching because of $w: C \cond D$, we can permute
($\cond$ L) over the other rules, since $w$ is a predecessor of
$x$ in the tree of transitions (see the definition \ref{sequente
x-branching} above): the label used to decompose the conditional
formula is already in the sequent, then the permutation is
possible\footnote{As explained, ($\cond$ L) does not permute over
the application of ($\cond$ R) which introduces the label used by
($\cond$ L).}. Then we can apply the disjunction property and
obtain that for each $i$,
\begin{center}
  $\Gamma_i \vdash \Delta_i, x: A \Ri B$ is derivable.
\end{center}
Thus, deleting one occurrence of $x: A \Ri B$ in the consequent of
any sequent in $\Pi'$ below $\Gamma_i \vdash \Delta_i, x: A \Ri B,
x: A \Ri B$ we get a derivation of $\Gamma \vdash \Delta, x: A \Ri
B$. \finedim

\subsection{Elimination of Contractions in SeqCK and SeqID} In this subsection we show that we can
eliminate the application of contraction rules in SeqCK and SeqID
systems.

\begin{teorema}[Elimination of contractions in SeqCK and SeqID]
  Given a sequent $\prova x_0: D$, derivable in SeqCK or in SeqID, it has
  a derivation with no applications of (Contr L) and (Contr R).
\end{teorema}
\emph{Proof} (Sketch). As we proved above, we do not have to
consider the case of (Contr R) applied to conditional formulas and
contractions on transitions. The proof proceeds similarly to the
one of Theorem 9.1.1 in \cite{vigano}
 by
triple induction respectively: (i) on the number of contractions
in a proof of the sequent, (ii) on the complexity of the formula
involved in a contraction step, and (iii)  on the rank of the
contraction; we need the following two definitions:

\begin{definition}[Complexity of a formula cp($F$)]
  We define the complexity of a formula $F$ as follows:
    \begin{enumerate}
      \item cp $(x: A)$ = 2*$\tc$ $A$ $\tc$
      \item cp $(x \trans{A} y)$ = 2*$\tc$ $A$ $\tc$+1
    \end{enumerate}
  where $\tc$ $A$ $\tc$ is the number of symbols occurring in the
  string representing the formula $A$.
\end{definition}

\begin{definition}[Rank of a contraction]
We define the rank of a contraction as the largest number of steps
between the conclusions of a contraction and an upward sequent
containing at least one of the two copies of the formula that is
contracted.
\end{definition}
For example, given the following proof tree:
\[
  \begin{prooftree}
    \[
       \[
            \shortstack{ $\Pi$\\
           $\Gamma$, \emph{x: A, x: A} $\prova$ $\Delta, x: B, x: B$
           }
         \justifies \Gamma, x: A \prova \Delta, x: B, x: A \imp B \using (\imp R)
       \]
     \justifies \Gamma \prova \Delta, x: A \imp B, x: A \imp B \using (\imp R)
    \]
     \justifies \Gamma \prova \Delta, x: A \imp B  \using (Contr
     R)
  \end{prooftree}
\]
the rank of the contraction applied to $x: A$ $\imp$ $B$ is 2 (the
minimum rank available), since there are two sequents between the
conclusion of the (Contr R) rule and the first sequent in which
there are no instances of the constituent of the contraction.

The third induction, on the rank, is needed since the rule ($\Ri$
L), which does not permute over the ($\Ri$ R) rule: we cannot
assume that there is  a proof which introduces the two copies of
the conditional formulas by ($\Ri$ L) rule one after the other
(this happens when the introduction of the first copy is separated
by the introduction of the second copy by ($\Ri$ R) rule occurring
in the middle) and we need to consider separately the two cases.

To carry on the proof, suppose that a derivation $\Pi$ of $\vdash
x_0: D$ contains $i+1$ contractions. Concentrate on a maximal
instance of contraction, say on
  a formula $F$
(so that the portion of the derivation above this step is
contraction-free).  In order to eliminate this contraction step,
thereby obtaining a proof $\Pi'$ that contain $i$ contractions, we
proceed by induction on the complexity of $F$, and then by
induction on the rank of the contraction step. We only sketch the
proof of the most difficult case, the one of a left contraction on
a conditional formula $x: A \cond B$; the other cases are easy and
left to the reader. One can find the entire proof in \cite{posu}.

We consider proof trees where ($\cond$ L) is applied as follows:
\[
  \begin{prooftree}
    x \trans{A^{'}} y \prova x \trans{A} y \quad \quad \Gamma, y:
    B \prova \Delta
    \justifies \Gamma, x: A \cond B \prova \Delta \using (\cond L)
  \end{prooftree}
\]
In fact, no rules in SeqCK and SeqID introduces (looking forward)
a transition formula $x \trans{A} x$ in the consequent of a
sequent, then we can also apply the Theorem \ref{Proprieta di cond
L} to all the applications of ($\cond$ L).

\noindent Given the following proof:
\[
  \begin{prooftree}
    \shortstack{
      $\Pi_1$\\
      $\Gamma$, \emph{x: A $\cond$ B, x: A $\cond$ B} $\prova$ $\Delta$
    }
    \justifies \shortstack{
      $\Gamma$, \emph{x: A $\cond$ B} $\prova$ $\Delta$\\
      $\Pi_0$\\
      $\prova$ \emph{$x_0$: D}
    }
    \using (Contr L)
  \end{prooftree}
\]
we can obtain a proof $\Pi^{*}$ of the sequent $\Gamma$, \emph{x:
A $\cond$ B} $\prova$ $\Delta$, removing that contraction on $x: A
\cond B$.\\By induction on the rank of the contraction, we have
the following cases:
\begin{enumerate}
  \item \emph{Base: rank=2}: we have the following proof:

\begin{footnotesize}
\[
  \begin{prooftree}
    \[
      \[
        \[
          \shortstack{
          $\Pi_A$\\ \emph{x $\trans{A^{'}}$ y} $\prova$ \emph{x $\trans{A}$ y}
          }
          \[
            \shortstack{
          $\Pi_B$\\ \emph{x $\trans{A^{''}}$ w} $\prova$ \emph{x $\trans{A}$ w}
            }
            \quad
            \shortstack{
              $\Pi_C^*$\\ \emph{x $\trans{A^{'}}$ y}, \emph{x $\trans{A^{''}}$
              w}, $\Gamma_1$, \emph{y: B, w: B} $\prova$
              $\Delta_1$, \emph{y: $B^{'}$, w: $B^{''}$}
            }
             \justifies \Gamma_1, x: A \cond B, y: B, x \trans{A^{'}} y, x
             \trans{A^{''}} w \prova \Delta_1, y: B^{'}, w: B^{''}
              \using (\cond L)
          \]
           \justifies \Gamma_1, x: A \cond B, x: A \cond B, x \trans{A^{'}} y, x
           \trans{A^{''}} w \prova \Delta_1, y: B^{'}, w: B^{''}
            \using (\cond L)
        \]
         \justifies \Gamma_1, x: A \cond B, x \trans{A^{'}} y, x
         \trans{A^{''}} w \prova \Delta_1, y: B^{'}, w: B^{''}
         \using (Contr L)
      \]
      \justifies \Gamma_1, x: A \cond B, x \trans{A^{'}} y \prova
      \Delta_1, y: B^{'}, x: A^{''} \cond B^{''} \using (\cond R)
    \]
    \justifies \shortstack{
      $\Gamma_1$, \emph{x: A $\cond$ B} $\prova$ $\Delta_1$,
      \emph{x: $A^{'}$ $\cond$ $B^{'}$, x: $A^{''}$ $\cond$ $B^{''}$}\\
      $\Pi_0^*$\\
      $\prova$ \emph{$x_0$: D}
    }
    \using (\cond R)
  \end{prooftree}
\]
\end{footnotesize}
Notice that in $\Pi_A$, $\Pi_B$ and $\Pi_C^{*}$ there are no
applications of (Contr L) on $x: A \cond B$. If \emph{x
$\trans{A^{'}}$ y}, \emph{x $\trans{A^{''}}$ w}, $\Gamma_1$,
\emph{y: B, w: B} $\prova$ $\Delta_1, y: B^{'}, w: B^{''}$ is not
$x$-branching, then we can apply the Proposition \ref{proposizione
1 della tesi} to \emph{x $\trans{A^{'}}$ y}, \emph{x
$\trans{A^{''}}$ w}, $\Gamma_1$, \emph{y: B, w: B} $\prova$
$\Delta_1$, \emph{y: $B^{'}$, w: $B^{''}$}, obtaining that one of
the following sequents is derivable:
      \begin{enumerate}
        \item ($\Gamma_1$, \emph{x $\trans{A^{'}}$ y, x $\trans{A^{''}}$ w, y: B, w:
        B}$)^{*}_y$ $\prova$ $\Delta_{1y}^{*}$, \emph{y: $B^{'}$}
        \item ($\Gamma_1$, \emph{x $\trans{A^{'}}$ y, x $\trans{A^{''}}$ w, y: B, w:
        B}$)^{*}_w$ $\prova$ $\Delta_{1w}^{*}$, \emph{w: $B^{''}$}
        \item ($\Gamma_1$, \emph{x $\trans{A^{'}}$ y, x $\trans{A^{''}}$ w, y: B, w:
        B}) - (($\Gamma_1$, \emph{x $\trans{A^{'}}$ y, x $\trans{A^{''}}$ w, y: B, w:
        B}$)^{\pr}_y$ $\unione$ ($\Gamma_1$, \emph{x $\trans{A^{'}}$ y, x $\trans{A^{''}}$ w, y: B, w:
        B}$)^{\pr}_w$) $\prova$ $\Delta_1$ - ($\Delta_{1y}^{\pr}$ $\unione$ $\Delta_{1w}^{\pr}$)
      \end{enumerate}
We observe that \emph{x $\trans{A^{'}}$ y} and \emph{y: B} are
both members of ($\Gamma_1$, \emph{x $\trans{A^{'}}$ y, x
$\trans{A^{''}}$ w, y: B, w:
        B}$)^{*}_y$, whereas they are \emph{not} members of the
        multiset ($\Gamma_1$, \emph{x $\trans{A^{'}}$ y, x
$\trans{A^{''}}$ w, y: B, w:
        B}$)^{*}_w$; vice versa for the formulas \emph{x $\trans{A^{''}}$ w} and \emph{w:
B}. Then we have that one of the following sequents is derivable:
      \begin{enumerate}
        \item $\Gamma_{1y}^{*}, x \trans{A^{'}} y, y: B
        \prova \Delta_{1y}^{*}, y: B^{'}$
        \item $\Gamma_{1w}^{*}, x \trans{A^{''}} w, w: B
        \prova \Delta_{1w}^{*}, w: B^{''}$
        \item $\Gamma_1$ - ($\Gamma_{1y}^{\pr}$ $\unione$
        $\Gamma_{1w}^{\pr}$) $\prova$ $\Delta_1$ - ($\Delta_{1y}^{\pr}$ $\unione$ $\Delta_{1w}^{\pr}$)
      \end{enumerate}
In each of these cases, we can obtain a proof without adding any
contraction:
     \begin{enumerate}
        \item we have the following proof:\\
        \[
          \begin{prooftree}
        \[
         \[
          \[
            \shortstack{
              $\Pi_A$\\ \emph{x $\trans{A^{'}}$ y} $\prova$ \emph{x $\trans{A}$ y}
            }
            \quad
            \[
              \shortstack{
                $\Pi^{\pr}$ \\
                $\Gamma_{1y}^{*}$, \emph{x $\trans{A^{'}}$ y, y: B}
                $\prova$ $\Delta_{1y}^{*}$, \emph{y: $B^{'}$}
              }
              \justifies \Gamma_1, x \trans{A^{'}} y, y: B \prova \Delta_1, y: B^{'} \using (Weak)
            \]
            \justifies \Gamma_1, x \trans{A^{'}} y, x: A \cond B
            \prova \Delta_1, y: B^{'} \using (\cond L)
          \]
          \justifies \Gamma_1, x \trans{A^{'}} y, x \trans{A^{''}}
          w, x: A \cond B \prova \Delta_1, y: B^{'}, w: B^{''}
          \using (Weak)
        \]
      \justifies \Gamma_1, x: A \cond B, x \trans{A^{'}} y \prova
      \Delta_1, y: B^{'}, x: A^{''} \cond B^{''} \using (\cond R)
    \]
    \justifies \shortstack{
      $\Gamma_1$, \emph{x: A $\cond$ B} $\prova$ $\Delta_1$,
      \emph{x: $A^{'}$ $\cond$ $B^{'}$, x: $A^{''}$ $\cond$ $B^{''}$}\\
      $\Pi_0^*$\\
       $\prova$ \emph{$x_0$: D}
         }
           \using (\cond R)
          \end{prooftree}
        \]

      \item symmetric to the previous case;

       \item in this case, we obtain:\\
       \[
         \begin{prooftree}
         \[
         \shortstack{
           $\Pi^{\pr}$\\
$\Gamma_1$ - ($\Gamma_{1y}^{\pr}$ $\unione$
        $\Gamma_{1w}^{\pr}$) $\prova$ $\Delta_1$ - ($\Delta_{1y}^{\pr}$ $\unione$ $\Delta_{1w}^{\pr}$)
         }
         \justifies \Gamma_1 \prova \Delta_1
         \using (Weak)
         \]
    \justifies \shortstack{
      $\Gamma_1$, \emph{x: A $\cond$ B} $\prova$ $\Delta_1$,
      \emph{x: $A^{'}$ $\cond$ $B^{'}$, x: $A^{''}$ $\cond$ $B^{''}$}\\
      $\Pi_0^*$\\
       $\prova$ \emph{$x_0$: D}
         }
           \using (Weak)
          \end{prooftree}
       \]
      \end{enumerate}
If \emph{x $\trans{A^{'}}$ y}, \emph{x $\trans{A^{''}}$
              y}, $\Gamma_1$, \emph{y: B, w: B} $\prova$
              $\Delta_1$, \emph{y: $B^{'}$, w: $B^{''}$} is
$x$-branching, we can permute all the rules introducing
$x$-branching formulas over the others of the subtree $\Pi_C^{*}$,
in a similar way to the proof of the disjunction property. We then
apply the Proposition \ref{proposizione 1 della tesi} to the non
$x$-branching sequents $\Gamma_{1i}, x \trans{A^{'}} y, x
\trans{A^{''}} w, y: B, w: B \prova \Delta_{1i}, y: B^{'}, w:
B^{''}$, obtaining a derivation for one of the following:
\begin{enumerate}
        \item $\Gamma_{1iy}^{*}$, \emph{x $\trans{A^{'}}$ y, y: B}
        $\prova$ $\Delta_{1iy}^{*}$, \emph{y: $B^{'}$}
        \item $\Gamma_{1iw}^{*}$, \emph{x $\trans{A^{''}}$ w, w: B}
        $\prova$ $\Delta_{1iw}^{*}$, \emph{w: $B^{''}$}
        \item $\Gamma_{1i}$ - ($\Gamma_{1iy}^{\pr}$ $\unione$
        $\Gamma_{1iw}^{\pr}$) $\prova$ $\Delta_{1i}$ - ($\Delta_{1iy}^{\pr}$ $\unione$ $\Delta_{1iw}^{\pr}$)
\end{enumerate}
In each case (proceeding as in the previous case) we can obtain a
proof of the sequents $\Gamma_{1i}, x \trans{A^{'}} y, x
\trans{A^{''}} w, x: A \cond B \prova \Delta_{1i}, y: B^{'}, w:
B^{''}$. Be $\Pi^{**}_C$ the proof obtained by permuting the rules
introducing $x$-branching formulas in $\Pi^*_C$; reapplying all
the rules of $\Pi^{**}_C$ to the sequents \emph{$\Gamma_{1i}$, x
$\trans{A^{'}}$ y,
      x $\trans{A^{''}}$ w, x: A
      $\cond$ B $\prova$ $\Delta_{1i}$, y: $B^{'}$, w: $B^{''}$}, removing from
the antecedent of each sequent an instance of $y: B$ and $w: B$
and adding an occurrence of $x: A \cond B$, we have a proof
$\Pi^{\pr\pr}$ of the sequent:
     \[
        \Gamma_1, x \trans{A^{'}} y, x \trans{A^{''}} w, x: A \cond B
        \prova \Delta_1, y: B^{'}, w: B^{''}
      \]
  from which we can obtain:
   \[
      \begin{prooftree}
        \[
        \shortstack{
          $\Pi^{\pr\pr}$\\
        \emph{$\Gamma_1$, x $\trans{A^{'}}$ y, x $\trans{A^{''}}$ w, x: A $\cond$ B
        $\prova$ $\Delta_1$, y: $B^{'}$, w: $B^{''}$}
        }
          \justifies
        \Gamma_1, x \trans{A^{'}} y, x: A \cond B
        \prova \Delta_1, y: B^{'}, x: A^{''} \cond B^{''}
        \using (\cond R)
        \]
        \justifies
        \shortstack{
          \emph{$\Gamma_1$, x: A $\cond$ B} $\prova$ $\Delta_1$,
          \emph{x: $A^{'}$ $\cond$ $B^{'}$, x: $A^{''}$ $\cond$
          $B^{''}$}\\ $\Pi_0$\\
          \emph{$\prova$ $x_0$: D}
        }
        \using (\cond R)
      \end{prooftree}
    \]

  \item \emph{Inductive step: rank\texttt{>}2}: the situation is
  as follows:
\[
  \begin{prooftree}
       \[
       \shortstack{
         $\Pi_A$\\ \emph{x $\trans{A^{'}}$ y $\prova$ x $\trans{A}$ y}
       }
       \[
         \shortstack{
           $\Pi_B$\\
           \emph{x $\trans{A^{''}}$ w $\prova$ x $\trans{A}$ w}
         }
         \quad
         \shortstack{
           $\Pi_E$\\
           \emph{x $\trans{A^{''}}$ w}, $\Gamma_2$, \emph{w: B}
           $\prova$ $\Delta_2$
         }
         \justifies \shortstack{
           \emph{x $\trans{A^{''}}$ w}, $\Gamma_2$, \emph{x: A $\cond$ B}
           $\prova$ $\Delta_2$\\
           $\Pi_D$\\
           \emph{x $\trans{A^{'}}$ y, $\Gamma_1$, y: B, x: A $\cond$ B
           $\prova$} $\Delta_1$
         } \using (\cond L)
       \]
         \justifies x \trans{A^{'}} y, \Gamma_1, x: A \cond B, x:
         A \cond B \prova \Delta_1 \using (\cond L)
       \]
        \justifies
        \shortstack{
          \emph{x $\trans{A^{'}}$ y}, $\Gamma_1$, \emph{x: A $\cond$ B} $\prova$ $\Delta_1$\\
          $\Pi_0$\\
          \emph{$\prova$ $x_0$: D}
        }
        \using (Contr L)
  \end{prooftree}
\]
We proceed in order to reduce the rank of the contraction and then
apply the inductive hypothesis; we have the following two
subcases:\\(1) \emph{x $\trans{A^{''}}$ w} $\appartiene$
  $\Gamma_1$: in this case $\Pi_D$ is not empty,
  (otherwise rank=2). we can permute the upper ($\cond$ L)
  over the rules in $\Pi_D$, reducing the
  rank;\\ (2) the transition formula \emph{x $\trans{A^{''}}$ w} is introduced by ($\cond$ R) in $\Pi_D$, then we have a
  proof like:\\
\[
  \begin{prooftree}
       \[
       \shortstack{
         $\Pi_A$\\ \emph{x $\trans{A^{'}}$ y $\prova$ x $\trans{A}$ y}
       }
       \[
           \[
         \shortstack{
           $\Pi_B$\\
           \emph{x $\trans{A^{''}}$ w $\prova$ x $\trans{A}$ w}
         }
         \quad
         \shortstack{
           $\Pi_E$\\
           \emph{x $\trans{A^{''}}$ w}, $\Gamma_2$, \emph{w: B}
           $\prova$ $\Delta_2$
         }
             \justifies \shortstack{
               \emph{x $\trans{A^{''}}$ w}, $\Gamma_2$, \emph{x: A $\cond$
               B} $\prova$ $\Delta_2$\\
               $\Pi_D^{''}$\\
               \emph{x $\trans{A^{''}}$ w}, $\Gamma_2^{'}$, \emph{x: A $\cond$ B}
               $\prova$ $\Delta_2^{'}$, \emph{w: $B^{''}$}
             }
             \using (\cond L)
           \]
         \justifies \shortstack{
           $\Gamma_2^{'}$, \emph{x: A $\cond$ B}
           $\prova$ $\Delta_2^{'}$, \emph{x: $A^{''}$ $\cond$ $B^{''}$}\\
           $\Pi_D^{'}$\\
           \emph{x $\trans{A^{'}}$ y, $\Gamma_1$, y: B, x: A $\cond$ B
           $\prova$} $\Delta_1$
         } \using (\cond R)
         \]
         \justifies x \trans{A^{'}} y, \Gamma_1, x: A \cond B, x:
         A \cond B \prova \Delta_1 \using (\cond L)
       \]
        \justifies
        \shortstack{
          \emph{x $\trans{A^{'}}$ y}, $\Gamma_1$, \emph{x: A $\cond$ B} $\prova$ $\Delta_1$\\
          $\Pi_0$\\
          \emph{$\prova$ $x_0$: D}
        }
        \using (Contr L)
  \end{prooftree}
\]
$\Pi_D^{''}$ could be empty: if it is not, we can easily diminish
the rank of the contraction by permuting the upper ($\cond$ L)
over the rules of $\Pi_D^{''}$; then, we consider the most
difficult case that $\Pi_D^{''}$ is empty. We observe that:
  \begin{enumerate}
    \item \emph{x, y} and \emph{w} are all distinct;
    \item \emph{x: $A^{''}$ $\cond$ $B^{''}$} is not a subformula of \emph{y: B}, as \emph{x} is a
    predecessor of \emph{y}; it is necessarily a subformula of a formula in $\Gamma_1$ or in $\Delta_1$.
  \end{enumerate}
We can divide the subtree $\Pi_D^{'}$ in two subproofs,
$\Pi_D^{'a}$ and $\Pi_D^{'b}$, such that $\Pi_D^{'b}$ introduces
\emph{y: B} and \emph{x $\trans{A^{'}}$ y}\footnote{If
$\Pi_D^{'b}$ is empty, we have that \emph{y: B} and \emph{x
$\trans{A^{'}}$ y} are both in $\Gamma_1$.}, whereas the formula
\emph{x: $A^{''}$ $\cond$ $B^{''}$} is used as a premise of a rule
in $\Pi_D^{'a}$. Due this separation, we can permute ($\cond$ R)
over the lowest ($\cond$ L), obtaining the following proof:
\[
\begin{prooftree}
        \[
          \[
            \shortstack{
              $\Pi_A$\\
              \emph{x $\trans{A^{'}}$ y} $\prova$ \emph{x $\trans{A}$ y}
            }
            \quad
            \[
              \shortstack{
                $\Pi_B$\\
                \emph{x $\trans{A^{''}}$ w} $\prova$ \emph{x $\trans{A}$ w}
              }
              \quad \quad
              \shortstack{
                $\Pi_E$\\
                \emph{x $\trans{A^{''}}$ w}, $\Gamma_2$, \emph{y: B} $\prova$ $\Delta_2$
              }
              \justifies
              \shortstack{
              \emph{x $\trans{A^{''}}$ w}, $\Gamma_2$, \emph{x: A $\cond$
              B} $\prova$ $\Delta_2$\\
              $\Pi_D^{''}$\\
              \emph{x $\trans{A^{''}}$ w}, $\Gamma_2^{'}$, \emph{x: A $\cond$
              B} $\prova$ $\Delta_2^{'}$, \emph{w: $B^{''}$}\\
              $\Pi_D^{'b}$\\
              \emph{x $\trans{A^{'}}$ y}, \emph{x $\trans{A^{''}}$ w},
              $\Gamma_1^*$, \emph{y: B, x: A $\cond$ B} $\prova$
              $\Delta_2^*$, \emph{w: $B^{''}$}
              }
              \using (\cond L)
            \]
            \justifies x \trans{A^{'}} y, x \trans{A^{''}} w,\Gamma_1^*, x: A \cond
            B, x: A \cond B \prova \Delta_1^*, w: B^{''} \using
            (\cond L)
          \]
        \justifies
        \shortstack{
          \emph{x $\trans{A^{'}}$ y}, $\Gamma_1^*$, \emph{x: A $\cond$
          B}, \emph{x: A $\cond$ B} $\prova$ $\Delta_1^*$, \emph{x: $A^{''}$ $\cond$
          $B^{''}$}\\
          $\Pi_D^{'a}$\\
          \emph{x $\trans{A^{'}}$ y}, $\Gamma_1$, \emph{x: A $\cond$ B}, \emph{x: A $\cond$ B}
           $\prova$ $\Delta_1$
        }
         \using (\cond R)
        \]
        \justifies \shortstack{
          \emph{x $\trans{A^{'}}$ y}, $\Gamma_1$, \emph{x: A $\cond$ B} $\prova$ $\Delta_1$\\
          $\Pi_0$\\
          \emph{$\prova$ $x_0$: D}
        }
        \using (Contr L)
  \end{prooftree}
\]
Now we can permute ($\cond$ R) over the contraction rule, obtaining the tree:\\
\[
  \begin{prooftree}
    \[
      \[
            \shortstack{
              $\Pi_A$\\
              \emph{x $\trans{A^{'}}$ y} $\prova$ \emph{x $\trans{A}$ y}
            }
            \quad
            \[
              \shortstack{
                $\Pi_B$\\
                \emph{x $\trans{A^{''}}$ w} $\prova$ \emph{x $\trans{A}$ w}
              }
              \quad \quad
              \shortstack{
                $\Pi_E$\\
                \emph{x $\trans{A^{''}}$ w}, $\Gamma_2$, \emph{y: B} $\prova$ $\Delta_2$
              }
              \justifies
              \shortstack{
              \emph{x $\trans{A^{''}}$ w}, $\Gamma_2$, \emph{x: A $\cond$
              B} $\prova$ $\Delta_2$\\
              $\Pi_D^{''}$\\
              \emph{x $\trans{A^{''}}$ w}, $\Gamma_2^{'}$, \emph{x: A $\cond$
              B} $\prova$ $\Delta_2^{'}$, \emph{w: $B^{''}$}\\
              $\Pi_D^{'b}$\\
              \emph{x $\trans{A^{'}}$ y}, \emph{x $\trans{A^{''}}$ w},
              $\Gamma_1^*$, \emph{y: B, x: A $\cond$ B} $\prova$
              $\Delta_2^*$, \emph{w: $B^{''}$}
              }
          \using (\cond L)
        \]
        \justifies x \trans{A^{'}} y, x \trans{A^{''}} w,
      \Gamma_1^*, x: A \cond B, x: A \cond B \prova w: B^{''}, \Delta_1^*
      \using(\cond L)
      \]
      \justifies x \trans{A^{'}} y, x \trans{A^{''}} w,
      \Gamma_1^*, x: A \cond B \prova w: B^{''}, \Delta_1^* \using
      (Contr L)
    \]
    \justifies \shortstack{
      \emph{x $\trans{A^{'}}$ y}, $\Gamma_1^*$, \emph{x: A $\cond$
      B} $\prova$ $\Delta_1^*$, \emph{x: $A^{''}$ $\cond$
      $B^{''}$}\\
      $\Pi_D^{'a*}$\\
      \emph{x $\trans{A^{'}}$ y}, $\Gamma_1$, \emph{x: A $\cond$
      B} $\prova$ $\Delta$\\
      $\Pi_0$\\
      $\prova$ \emph{$x_0$: D}
    }
    \using (\cond R)
  \end{prooftree}
\]
in which the subtree $\Pi^{'a*}_D$ is obtained by deleting an
occurrence of $x: A \cond B$ in every sequent descending from
\emph{x $\trans{A^{'}}$ y, $\Gamma_1^*$,} $x: A \cond B, x: A
\cond B$ $\prova$ $\Delta_1^*$, \emph{x: $A^{''}$ $\cond$
$B^{''}$} in $\Pi^{'a}_D$.\\In this way, the rank of the
contraction is diminished and we can apply the inductive
hypothesis.

\end{enumerate}
\finedim

\subsection{Bound for the Application of (Contr L) in SeqMP and SeqID+MP}
In SeqMP and SeqID+MP we cannot completely eliminate the
application of the left contraction rule on conditional formulas
$x: A \cond B$; for example, the following sequent:
\begin{center}
$x: \vero \cond (B \andd \neg(\vero \cond B)) \prova$
\end{center}
 is valid in CK+MP, but it can only be derived in SeqMP by applying
the (Contr L) rule on the conditional formula $x: \vero \cond (B
\andd \neg (\vero \cond B))$, as follows:
\begin{footnotesize}
\[
  \begin{prooftree}
    \[
      \[
        ... \prova x: \vero
        \justifies ... \prova x \trans{\vero} x \using (MP)
      \]
      \[
        \[
          \[
            \[
                x \trans{\vero} y,...\prova x \trans{\vero} y,...
                \quad \quad
                \[
                y: B, y: \neg (\vero \cond B),... \prova y: B
                \justifies y: B \andd \neg (\vero \cond B),... \prova y:
                B \using (\andd L)
                \]
              \justifies x: B, x \trans{\vero} y, x: \vero \cond (B \andd \neg (\vero \cond
      B)) \prova y: B \using (\cond L)
            \]
            \justifies x: B, x: \vero \cond (B \andd \neg (\vero \cond
      B)) \prova x: \vero \cond B \using (\cond R)
          \]
          \justifies x: B, x: \neg (\vero \cond B), x: \vero \cond (B \andd \neg (\vero \cond
      B)) \prova \using (\neg L)
        \]
        \justifies x: B \andd \neg (\vero \cond B), x: \vero \cond (B \andd \neg (\vero \cond
      B)) \prova \using (\andd L)
      \]
      \justifies x: \vero \cond (B \andd \neg (\vero \cond B)), x: \vero \cond (B \andd \neg (\vero \cond
      B)) \prova \using (\cond L)
    \]
    \justifies x: \vero \cond (B \andd \neg (\vero \cond B))
    \prova \using (Contr L)
  \end{prooftree}
\]
\end{footnotesize}

\noindent The sequent is not derivable without any application of
(Contr L) on conditionals, as shown by the following tree:

\[
  \begin{prooftree}
    \[
      \prova x: \vero
    \justifies \prova x \trans{\vero} x \using (MP)
    \]
    \[
      \[
        \[
          x: B, x \trans{\vero} y \prova y: B
          \justifies x: B \prova x: \vero \cond B \using (\cond R)
        \]
        \justifies x: B, x: \neg (\vero \cond B) \prova \using (\neg L)
      \]
      \justifies x: B \andd \neg (\vero \cond B) \prova \using
      (\andd L)
    \]
    \justifies x: \vero \cond (B \andd \neg (\vero \cond B))
    \prova \using (\cond L)
  \end{prooftree}
\]

\noindent $x: B, x \trans{\vero} y \prova y: B$, where $B$ is an
atom, cannot be proved.

In order to obtain a decision procedure for this systems, we must
control the application of the (Contr L) rule. We show that it is
sufficient to apply (Contr L) at most once on each conditional
formula $x: A \cond B$ in every branch of a proof tree; we say
that contractions on non-conditional formulas and multiple
contractions (more than one) on conditionals are \emph{redundant};
we show that we can eliminate all the redundant contractions.

\begin{definition}[Multiple contractions]
  Given a proof tree $\Pi$, we say that one branch $\BB$ of $\Pi$ has
  multiple contractions if a contraction rule is applied to a
  formula $F$ $n$ times, $n$ \texttt{>} 1.
\end{definition}
\begin{definition}[Redundant contractions on a formula $F$]
  Given a proof tree $\Pi$, we say that it has
  redundant contractions on a formula $F$ if it has a branch $\BB$ with at least one of the following conditions:
  \begin{enumerate}
    \item a contraction rule is applied to a \emph{non}
    conditional formula $F$;
    \item the right contraction rule (Contr R) is applied to a
    conditional formula $F$;
    \item there are multiple contractions of (Contr L) on a
    conditional formula $F$.
  \end{enumerate}
\end{definition}
\begin{definition}[integer multiset ordering \texttt{$<_m$}]
  Given $\Gamma$ and $\Delta$, multisets of integers, we say that:
\begin{enumerate}
  \item $\Gamma$ \texttt{$<_m$} $\Delta$ if $\Gamma$ $\subset$
  $\Delta$;
  \item $\Gamma$ \texttt{$<_m$} $\Delta$ if $\Gamma$ \texttt{$<_m$}
  $\Delta^{'}$, where $\Delta^{'}$ = $\Delta$ - \{ $j$ \} $\unione$
  \{ $i, i, ..., i$ \} and $i$ \texttt{<} $j$.
\end{enumerate}
\end{definition}
As it is well known, $<_m$ is a well-order on multisets.

In the proof for bounding contractions in SeqMP and SeqID+MP we
need the following:

\begin{lemmaPosu}\label{lemma 2 tesi Posu}
If $\Gamma, x: A \cond B, x: A \cond B \prova \Delta, x: A$ has a
derivation $\Pi$ with no contractions on $x: A \cond B$ in SeqMP
(SeqID+MP), then $\Gamma, x: A \cond B \prova \Delta, x: A$ is
derivable in SeqMP (SeqID+MP) and it has a derivation which at
most adds to the contractions in $\Pi$ contractions on formulas
with lower complexities than the complexity of $x: A \cond B$.
\end{lemmaPosu}

\noindent \emph{Proof}. If an instance of $x: A \cond B$ is
introduced in $\Pi$ by implicit weakening, then we can conclude by
deleting this weakening. If the two instances of $x: A \cond B$
are both derived from ($\cond$ L) using transitions of the form $x
\trans{A} y$, with $x \diverso y$, then the contraction can be
eliminated as in CK\{+ID\}. If an instance of $x: A \cond B$ is
introduced in $\Pi$ by an application of ($\cond$ L) using a
transition $x \trans{A} x$ derived from (MP), then it permutes
over the other rules in $\Pi$; therefore, we have the following
proof:
\[
\begin{prooftree}
  \[
    \shortstack{$\Pi_A$ \\ $\Gamma, x: A \cond B \prova \Delta, x: A, x: A$}
    \justifies \Gamma, x: A \cond B \prova \Delta, x: A, x
    \trans{A} x \using (MP)
  \]
  \shortstack{$\Pi_B$ \\ $\Gamma, x: A \cond B, x: B \prova \Delta, x: A$}
  \justifies \Gamma, x: A \cond B, x: A \cond B \prova \Delta, x:
  A \using (\cond L)
\end{prooftree}
\]
We can conclude the proof by adding a contraction on the
sub-formula $x: A$ as follows:
\[
\begin{prooftree}
  \shortstack{$\Pi_A$ \\ $\Gamma, x: A \cond B \prova \Delta, x: A, x: A$}
  \justifies \Gamma, x: A \cond B \prova \Delta, x: A \using
  (Contr R)
\end{prooftree}
\]
\finedim

\begin{definition}[c($\Pi$)]
Given a proof tree $\Pi$ we define c($\Pi$) as the multiset of
integers of the complexities of the formulas to which are applied
redundant contractions in $\Pi$.
\end{definition}
For example, consider a proof tree $\Pi$ having three branches
with the following features: a contraction on a formula $x: A \imp
B$ with complexity 7 and two applications of (Contr L) on $x: C
\cond D$ with complexity 5 in the left branch; an application of
(Contr L) on a conditional formula $x: E \cond F$ with complexity
9 in the central branch; two contractions on an atomic formula
(complexity 2) and three contractions (Contr L) on $x: G \cond H$
with complexity 5 in the right branch. We have that c($\Pi$)=\{7,
5, 2, 5\} (the contraction in the central branch is not
redundant).

We can control the application of the contractions in SeqMP and
SeqID+MP as explained by the following:
\begin{teorema}[Bound for the contractions in SeqMP and SeqID+MP]
  Given a sequent $\prova x_0: D$, derivable in SeqMP or in SeqID+MP,
  then it has a derivation where there is at most one application
  of (Contr L) on each conditional formula $x: A \cond B$ in every branch of the proof tree.
  In other words, we can find a derivation of it with no redundant contractions.
\end{teorema}

\noindent \emph{Proof} (Sketch). Using the results of the previous
subsections, we can say, without loss of generality, that the
sequent $\prova x_0: D$ has a derivation $\Pi$ with no
applications of (Contr R) on conditional formulas and no
contractions on transitions. Then, we proceed by induction on the
multiset ordering defined by c($\Pi$). The idea is the following:
at each step of the proof, we build a proof tree $\Pi^{'}$ such
that c($\Pi^{'}$) \texttt{$<_m$} c($\Pi$), to which we apply the
inductive hypothesis; $\Pi^{'}$ is obtained by removing the
redundant contraction on a formula with complexity $M$ such that
$M=max$(c($\Pi$)). We can have two different situations:
\begin{enumerate}
  \item the redundant contractions on formulas with complexity
  $M$ are deleted \emph{without adding any other contraction in the proof
  tree}; therefore, we have that c($\Pi^{'}$)=c($\Pi$) - \{ \emph{M} \};
  \item the redundant contractions on formulas with complexity
  $M$ are replaced by redundant contractions on formulas with
  complexities \emph{$M_1$, $M_2$, ..., $M_k$}, where \emph{$M_i$} \texttt{<}
  \emph{M}. c($\Pi^{'}$) is then obtained by replacing an occurrence
  of $M$ in c($\Pi$) with the values $M_1$, $M_2$, ...,
  $M_k$.
\end{enumerate}
We only present the most interesting case; the entire proof is
contained in \cite{posu}, pages 163-198.\\Consider a branch $\BB$
with $n$ applications of (Contr L), $n$ \texttt{>} 1, on $x: A
\cond B$, where $max$(c($\Pi$)) = cp(\emph{x: A $\cond$ B}); we
can consider that all the $n$ contractions are applied in
sequence, then we consider the upper two instances of (Contr L)
and proceed to eliminate the upper one:
\[
  \begin{prooftree}
    \[
      \shortstack{
        $\Pi$\\
        $\Gamma$, \emph{x: A $\cond$ B}, \emph{x: A $\cond$ B}, \emph{x: A $\cond$ B} $\prova$ $\Delta$
      }
      \justifies \Gamma, x: A \cond B, x: A \cond B \prova \Delta
      \using (Contr L)
    \]
    \justifies
      \shortstack{
        $\Gamma$, \emph{x: A $\cond$ B} $\prova$ $\Delta$\\
        $\Pi_0$\\ $\prova$ \emph{$x_0$: D}
      }
    \using (Contr L)
  \end{prooftree}
\]
$\Pi$ does not contain any application of (Contr L) with
constituent $x: A \cond B$.\\We can also observe that an instance
of $x: A \cond B$ is the principal formula of the sequent, by the
permutability of the rules\footnote{As we explained above,
($\cond$ L) does not always permute over the ($\cond$ R) rule. We
can assume, without loss of generality, that $x: A \cond B$ is the
principal formula since, if all its three instances are derived by
($\cond $ L) and the labels used are introduced by ($\cond$ R) in
$\Pi$, then we can permute ($\cond$ R) over the contractions;
therefore, ($\cond$ L) can be permuted over the rules of $\Pi$.}.
We present the following situation: the ($\cond$ L) rule is
applied to $\Gamma$, \emph{x: A $\cond$ B}, \emph{x: A $\cond$ B},
\emph{x: A $\cond$ B} $\prova$ $\Delta$ using the label $x$:
\[
  \begin{prooftree}
    \[
      \[
      \shortstack{
        $\Pi_1$\\
        $\Gamma$, \emph{x: A $\cond$ B}, \emph{x: A $\cond$ B} $\prova$
        $\Delta$, \emph{x $\trans{A}$ x}
      }
      \quad \quad
      \shortstack{
        $\Pi_2$\\
        $\Gamma$, \emph{x: A $\cond$ B}, \emph{x: A $\cond$ B}, \emph{x: B} $\prova$
        $\Delta$
      }
      \justifies \Gamma, x: A \cond B, x: A \cond B, x: A \cond B
      \prova \Delta \using (\cond L)
      \]
      \justifies \Gamma, x: A \cond B, x: A \cond B \prova \Delta
      \using (Contr L)
    \]
    \justifies
      \shortstack{
        $\Gamma$, \emph{x: A $\cond$ B} $\prova$ $\Delta$\\
        $\Pi_0$\\ $\prova$ \emph{$x_0$: D}
      }
    \using (Contr L)
  \end{prooftree}
\]

If the transition formula $x \trans{A} x$ is introduced by
implicit weakening, then we have that $\Gamma$, \emph{x: A $\cond$
B}, \emph{x: A $\cond$ B} $\prova$ $\Delta$ is derivable and we
can immediately conclude (the upper (Contr L) has been
eliminated); if $x \trans{A} x$ is not introduced by weakening, it
can only be treated (looking backward) by the (MP) rule, then we
consider the following proof:
\[
  \begin{prooftree}
    \[
      \[
        \[
      \shortstack{
        $\Pi_1^{'}$\\
        $\Gamma$, \emph{x: A $\cond$ B}, \emph{x: A $\cond$ B} $\prova$
        $\Delta$, \emph{x: A}
      }
      \justifies \Gamma, x: A \cond B, x: A \cond B \prova x
      \trans{A} x, \Delta \using (MP)
      \]
      \shortstack{
        $\Pi_2$\\
        $\Gamma$, \emph{x: A $\cond$ B}, \emph{x: A $\cond$ B}, \emph{x: B} $\prova$
        $\Delta$
      }
      \justifies \Gamma, x: A \cond B, x: A \cond B, x: A \cond B
      \prova \Delta \using (\cond L)
      \]
      \justifies \Gamma, x: A \cond B, x: A \cond B \prova \Delta
      \using (Contr L)
    \]
    \justifies
      \shortstack{
        $\Gamma$, \emph{x: A $\cond$ B} $\prova$ $\Delta$\\
        $\Pi_0$\\ $\prova$ \emph{$x_0$: D}
      }
    \using (Contr L)
  \end{prooftree}
\]
By Lemma \ref{lemma 2 tesi Posu}, $\Gamma$, \emph{x: A $\cond$ B}
$\prova$ $\Delta$, \emph{x: A} is derivable with a proof
$\Pi^{\pr}$ that does not add any contraction on $x: A \cond
B$; we can obtain:\\
\[
  \begin{prooftree}
    \shortstack{
      $\Pi^{\pr}$\\ $\Gamma$, \emph{x: A $\cond$ B} $\prova$
      $\Delta$, \emph{x: A}
    }
    \justifies \Gamma, x: A \cond B \prova \Delta, x \trans{A} x
    \using (MP)
  \end{prooftree}
\]
Our target is now to find a proof tree $\Pi_2^*$, with no
contractions on $x: A \cond B$, of the sequent $\Gamma, x: A \cond
B, x: B \prova \Delta$. Consider the case when both the
occurrences of $x: A \cond B$ are introduced in $\Pi_2$ by the
($\cond$ L) rule. We analyze the situation in which
   an occurrence of $x: A \cond B$ is introduced by ($\cond$
  L) by using the label $x$; we can then consider the following
  proof tree:
      \[
        \begin{prooftree}
          \[
            \shortstack{
              $\Pi_A$\\ $\Gamma$, \emph{x: A $\cond$ B}, \emph{x:
              B} $\prova$ $\Delta$, \emph{x: A}
            }
            \justifies \Gamma, x: A \cond B, x: B \prova \Delta,
            x \trans{A} x \using (MP)
          \]
          \shortstack{
            $\Pi_B$\\ $\Gamma$, \emph{x: A $\cond$ B}, \emph{x:
            B}, \emph{x: B} $\prova$ $\Delta$
          }
          \justifies \Gamma, x: A \cond B, x: A \cond B, x: B
          \prova \Delta \using (\cond L)
        \end{prooftree}
      \]
$\Pi_A$ and $\Pi_B$ do not contain any application of (Contr L) on
$x: A \cond B$, then we have the following proof:
      \[
        \begin{prooftree}
          \shortstack{
            $\Pi_B$\\ $\Gamma$, \emph{x: A $\cond$ B}, \emph{x: B, x:
            B} $\prova$ $\Delta$
          }
          \justifies \Gamma, x: A \cond B, x: B \prova \Delta
          \using (Contr L)
        \end{prooftree}
      \]
The other cases, when both the conditionals $x: A \cond B$ are
introduced without using the label $x$ in the applications of
($\cond$ L), are left to the reader.

We have found two proofs, with no contractions on \emph{x: A
$\cond$ B}, at most introducing contractions on formulas with
lower complexity than the complexity of \emph{x: A $\cond$ B}, of
the sequents $\Gamma$, \emph{x: A $\cond$ B} $\prova$ \emph{x
$\trans{A}$ x}, $\Delta$ and $\Gamma$, \emph{x: A $\cond$ B, x: B}
$\prova$ $\Delta$. We can then obtain the following proof, erasing
the upper application of (Contr L) in the initial proof tree :
\[
  \begin{prooftree}
    \[
    \[
    \shortstack{
      $\Pi^{\pr}$\\ $\Gamma$, \emph{x: A $\cond$ B} $\prova$
      $\Delta$, \emph{x: A}
    }
    \justifies \Gamma, x: A \cond B \prova \Delta, x \trans{A} x
    \using (MP)
    \]
      \quad \quad \quad \quad
      \shortstack{
        $\Pi_2^*$\\ $\Gamma$, \emph{x: A $\cond$ B}, \emph{x: B}
        $\prova$ $\Delta$
      }
      \justifies \Gamma, x: A \cond B, x: A \cond B \prova \Delta
      \using (\cond L)
    \]
    \justifies
      \shortstack{
        $\Gamma$, \emph{x: A $\cond$ B} $\prova$ $\Delta$\\
        $\Pi_0$\\ $\prova$ \emph{$x_0$: D}
      }
    \using (Contr L)
  \end{prooftree}
\]

\finedim

\subsection{Reformulation of SeqMP and SeqID+MP}
In the previous subsection we have found that, in SeqMP and
SeqID+MP, the left contraction rule (Contr L) is only needed if
applied to conditional formulas $x: A \cond B$ and, in particular,
\emph{at most once on each formula occurring in every derivation
branch of a proof tree}. By this fact, we can reformulate the
calculi for these two systems, obtaining {\bf BSeqMP} and {\bf
BSeqID+MP} (the prefix B stands for "bounded contractions") with
the following features:
\begin{enumerate}
  \item proof trees do not contain redundant contractions;
  \item contractions on $x: A \cond B$ are \emph{absorbed} into
  the ($\cond$ L) rule.
\end{enumerate}
First of all, we represent a single node of a proof tree as
\begin{center}
  $K \tc \Psi \tc \Gamma \prova \Delta$
\end{center}
$K$ is the set containing all the conditional formulas that have
already been contracted in that branch of the proof tree. The
antecedent of a sequent is then split into two parts:
\begin{enumerate}
  \item the set $\Psi$ of the conditional formulas duplicated by contraction;
  \item the multiset $\Gamma$ with the other formulas.
\end{enumerate}
 ($\cond$ L)
is split in three rules:
\begin{enumerate}
  \item ($\cond$ L$)_1$ is
  applied to $x: A \cond B \appartiene \Gamma$ if $x: A \cond B$ does not belong to $K$, i.e. if this conditional formula
  has not yet been contracted in that branch. The principal formula $x: A \cond B$ is
  decomposed and a copy of it is added to $\Psi$
  and to $K$;

  \item ($\cond$ L$)_2$ is
  applied to $x: A \cond B \appartiene \Gamma$ if $x: A \cond B$ belongs to $K$, i.e. it has already been contracted in
  that branch. The conditional formula $x: A \cond B$ is decomposed without
  adding any copy of it in the auxiliary sets $\Psi$
  and $K$;

  \item ($\cond$ L$)_3$ is
  applied to $x: A \cond B \appartiene \Psi$, i.e. $x: A \cond B$ has been previously duplicated by an application
  of ($\cond$ L$)_1$. The principal formula $x: A \cond B$ is decomposed without
  adding any copy of it in the auxiliary sets $\Psi$
  and $K$.
\end{enumerate}

\noindent In other words, if a conditional formula $x: A \cond B$
in $\Gamma$ has not been duplicated in a branch, then it is
decomposed by ($\cond$ L$)_1$, which adds a copy of it in $K$ and
in $\Psi$. $K$ keeps trace of duplicated formulas in that branch.
Duplicated conditionals in $\Psi$ will be only decomposed, but no
duplicated. If $x: A \cond B$ in $\Gamma$ has already been
duplicated in a branch, i.e. $x: A \cond B$ belongs to $K$, it is
only decomposed and no further duplicated.

Sequent calculi BSeqMP and BSeqID+MP are shown in Figure \ref{bseq}; we
omit the reformulation for axioms and
 some rules, since they are identical to SeqS's rules (see Figures \ref{Figura con SeqS} and
 \ref{Regole aggiuntive di SeqS}) with the exception of the form of the
 sequents.

\begin{figure}

\begin{footnotesize}
\fbox{ \(
\begin{array}{rl@{\quad}rl}
{\bf (\cond L)_1} & \irule{\shortstack{$K \unione \{x: A \cond B\}
\tc \Psi \unione \{x: A \cond B\} \tc \Gamma^{'} \prova  x
\trans{A} y, \Delta$\\ $K \unione \{x: A \cond B\} \tc \Psi
\unione \{x: A \cond B\} \tc y: B, \Gamma^{'} \prova \Delta$}} {K
\tc \Psi \tc x: A \cond B, \Gamma^{'} \prova \Delta} {\quad, if \quad x: A
\cond B \not\appartiene K} &
\\
\\
{\bf (\cond L)_2} & \irule{K \tc \Psi \tc \Gamma^{'} \prova x
\trans{A} y, \Delta \quad \quad \quad K \tc \Psi \tc y: B,
\Gamma^{'} \prova \Delta} {K \tc \Psi \tc x: A \cond B, \Gamma^{'}
\prova \Delta} {\quad, if \quad x: A \cond B \appartiene K} &
\\
\\
{\bf (\cond L)_3} & \irule{\shortstack{$K^{'} \unione \{ x: A
\cond B \} \tc \Psi^{'} \tc \Gamma \prova x \trans{A} y, \Delta$ \\
$K^{'} \unione \{ x: A \cond B \} \tc \Psi^{'} \tc y: B, \Gamma
\prova \Delta$}} {K^{'} \unione \{x: A \cond B\} \tc \Psi^{'}
\unione \{x: A \cond B\} \tc \Gamma \prova \Delta} {} &
\\
\\
{\bf (EQ)} & \irule{\vuoto \tc \vuoto \tc u: A \prova u: B \quad
\quad \quad \vuoto \tc \vuoto \tc u: B \prova u: A} {K \tc \Psi
\tc x \trans{A} y, \Gamma^{'} \prova x \trans{B} y, \Delta^{'}} {}
&
\\
\\
\end{array}
\) }
\end{footnotesize}
 \caption{Sequent calculi BSeqMP and BSeqID+MP.} \label{bseq}
\end{figure}

\noindent Next theorem follows immediately from the above
reformulation:

\begin{teorema}
  A sequent $\Gamma \prova \Delta$ is derivable in SeqMP
  (SeqID+MP) if and only if $\vuoto \tc \vuoto \tc \Gamma \prova
  \Delta$ is derivable in BSeqMP (BSeqID+MP).
\end{teorema}

We give a derivation in BSeqMP of the sequent $x: \vero \cond (B
\andd \nott (\vero \cond B)) \prova$, introduced at the beginning
of the previous subsection as an example of sequent derivable in
SeqMP with a necessary application of the left contraction on
conditionals:

\begin{footnotesize}
\[
\begin{prooftree}
  \shortstack{
  $\{x: \vero \cond (B \andd \nott (\vero \cond B))\} \tc \{x: \vero \cond (B \andd \nott (\vero \cond
  B))\} \tc \prova x \trans{\vero} x$\\
  $\{x: \vero \cond (B \andd \nott (\vero \cond B))\} \tc \{x: \vero \cond (B \andd \nott (\vero \cond
  B))\} \tc x: B \andd \nott(\vero \cond B) \prova$}
  \justifies \vuoto \tc \vuoto \tc x: \vero \cond (B \andd \nott
  (\vero \cond B)) \prova \using (\cond L)_1
\end{prooftree}
\]
\end{footnotesize}

\noindent The upper premise is derived as follows:

\[
\begin{prooftree}
\{x: \vero \cond (B \andd \nott (\vero \cond B))\} \tc \{x: \vero
\cond (B \andd \nott (\vero \cond
  B))\} \tc \prova x: \vero
  \justifies \{x: \vero \cond (B \andd \nott (\vero \cond B))\} \tc \{x: \vero \cond (B \andd \nott (\vero \cond
  B))\} \tc \prova x \trans{\vero} x \using (MP)
\end{prooftree}
\]

\noindent The other one has the following derivation:

\begin{footnotesize}
\[
\begin{prooftree}
  \[
    \[
      \[
        \[
         \{x: \vero \cond (B \andd \nott (\vero \cond B))\} \tc
        \vuoto \tc x \trans{\vero} y, x: B, y: B, y: \nott (\vero \cond
        B) \prova y: B
          \justifies \{x: \vero \cond (B \andd \nott (\vero \cond B))\} \tc
        \vuoto \tc x \trans{\vero} y, x: B, y: B \andd \nott (\vero \cond
        B) \prova y: B \using (\andd L)
        \]
        \quad \quad \Pi
        \justifies \{x: \vero \cond (B \andd \nott (\vero \cond B))\} \tc \{x: \vero \cond (B \andd \nott (\vero \cond
  B))\} \tc x \trans{\vero} y, x: B \prova y: B \using (\cond L)_3
      \]
      \justifies \{x: \vero \cond (B \andd \nott (\vero \cond B))\} \tc \{x: \vero \cond (B \andd \nott (\vero \cond
  B))\} \tc x: B \prova x: \vero \cond B \using (\cond R)
    \]
  \justifies \{x: \vero \cond (B \andd \nott (\vero \cond B))\} \tc \{x: \vero \cond (B \andd \nott (\vero \cond
  B))\} \tc x: B, x: \nott (\vero \cond B) \prova \using (\nott L)
  \]
  \justifies \{x: \vero \cond (B \andd \nott (\vero \cond B))\} \tc \{x: \vero \cond (B \andd \nott (\vero \cond
  B))\} \tc x: B \andd \nott(\vero \cond B) \prova \using (\andd
  L)
\end{prooftree}
\]
\end{footnotesize}

\noindent where $\Pi$ is the axiom $\{x: \vero \cond (B \andd
\nott (\vero \cond B))\} \tc \vuoto \tc x \trans{\vero} y, x: B
\prova x \trans{\vero} y, y: B$.

\subsection{Complexity of CK\{+ID\}}
Since we can eliminate contraction, it is relatively easy to prove
both decidability and a space  complexity bound.

In a proof without contractions, in  all rules the premises have a
smaller complexity than the conclusion. By this fact we get that
the length of each branch in a proof of a sequent $\vdash x_0: D$
is bounded by $O(\mid D\mid)$.

Moreover, observe that the rules are analytic, so that the
premises contains only (labelled) subformulas of the formulas in
the conclusion. In the  search of a proof of $\vdash x_0: D$, with
$\mid D\mid =n$, new labels are introduced only by (positive)
conditional subformulas of $D$. Thus, the  number of different
labels occurring in a proof is  $O(n)$; it follows that the total
number of   distinct labelled formulas is $O(n^2)$, and only
$O(n)$ of them can actually occur in each sequent.

This itself gives decidability:

\begin{teorema}[CK\{+ID\} decidability]
Logic CK\{+ID\} is decidable.
\end{teorema}

\noindent \emph{Proof}. We just observe that there is only a
finite number of derivations to check of a given sequent $\vdash
x_0: D$, as both the  length of a proof  and the number of
labelled formulas which may occur in it is finite. \finedim

Notice that, as usual, a proof  may have   an exponential size
since the branching introduced by the rules. However we can obtain
a much sharper  space complexity bound  using a standard technique
\cite{Hudelmaier,vigano}, namely we do not need to store the whole
proof, but only a sequent at a time plus additional information to
carry on the proof search. In searching a proof there are two
kinds of branching to consider:
 AND-branching caused by the rules with multiple premises and  OR-branching (backtracking points
in a depth first search) caused by the choice of the rule to
apply, and how to apply it in the case of ($\Ri$ L).

We store only one sequent at a time and maintain a stack
containing information sufficient to reconstruct the branching
points of both types. Each stack entry contains the principal
formula (either a labelled  sentence $x; B$, or a transition
formula $x\ff{B} y$), the name of the rule applied and an index
which allows to reconstruct the other branches on return to the
branching points.  The stack entries represent  thus backtracking
points and the index within the entry  allows one to reconstruct
both  the AND branching  and to check whether there are
alternatives to explore (OR branching). The working sequent on a
return point
 is recreated by replaying the stack entries from the bottom  of the stack using the information
in the index (for instance in the case of ($\Ri$ L) applied to the
principal formula $x: A\Ri B$, the index will indicate which
premise-first or second-we have to expand and the  label $y$
involved in the transition formula $x\ff{A} y$).

A proof begins with the end sequent $\vdash x_0: D$ and the empty
stack. Each rule application generates a new sequent and extends
the stack. If the current sequent is an axiom we pop the stack
until we find an AND branching point to be expanded. If there are
not, the end sequent $\vdash x_0: D$ is provable and we have
finished. If the current sequent is not an axiom and no rule can
be applied to it, we pop the stack entries and we continue at the
first available entry  with some  alternative  left (a
backtracking point). If there are no such entries, the end sequent
is not provable.

The entire process must terminate since: (i) the depth of the
stack is bounded by the length of a branch proof, thus it is
   $O(n)$, where $\mid D\mid = n$,  (ii)  the branching is
bounded by the number of rules, the number of  premises of any
rule and the number of formulas occurring in one sequent, the last
being  $O(n)$.

 To evaluate the space requirement, we have that each subformula of the initial formula can be
represented by a positional index into the  initial formula, which
requires $O(\log n)$ bits. Moreover, also each label can be
represented by $O(\log n)$ bits. Thus, to store the working
sequent we need $O(n~\log n)$ space, since there may occur  $O(n)$
labelled subformulas. Similarly, each stack entry requires $O(\log
n)$ bits,  as the name of the rule requires constant space and the
index $O(\log n)$ bits. Having depth $O(n)$, to store  the whole
stack requires $O(n~\log n)$ space. Thus we obtain:

\begin{teorema}[Space complexity of CK\{+ID\}]
Provability in CK\{+ID\} is decidable in $O(n~\log n)$ space.
\end{teorema}

\subsection{Decidability and Complexity of CK+MP\{+ID\}}
Let us conclude this section with a quick analysis of the systems
CK+MP\{+ID\}. We explained above that we cannot eliminate the
(Contr L) rule on conditional formulas in SeqMP and SeqID+MP
systems; however, we can control its application, in order to
obtain a decision procedure for these logics, too.

We can prove the following:
\begin{teorema}[CK+MP\{+ID\} decidability]
Logic CK+MP\{+ID\} is decidable.
\end{teorema}

\noindent \emph{Proof}. It is easy to prove that BSeqMP and
BSeqID+MP terminate, from which we obtain a decision procedure for
logics CK+MP and CK+MP+ID. As we explained above, the (Contr L)
rule is absorbed by the ($\cond$ L) rule; a conditional formula
$x: A \cond B$ is contracted at most one time on each branch with
an application of ($\cond$ L$)_1$ and (eventually) the following
application of ($\cond$ L$)_3$, therefore the length of each
branch in a proof is limited, as only a finite number of
conditional formulas can be introduced in that branch. \finedim

Let us say something about the space complexity of BSeqMP and
BSeqID+MP. These systems are decidable in exponential space in the
number of nested conditional formulas. Consider, for example, an
application of the ($\cond$ L$)_1$ rule with principal formula $x:
A_1 \cond (A_2 \cond B)$, which introduces the subformula $y: A_2
\cond B$ in the antecedent of one premise and is duplicated in
$K$; if the ($\cond$ L$)_1$ rule is also applied to $y: A_2 \cond
B$, then it is duplicated in $K$, too, obtaining two copies of $x:
A_1 \cond (A_2 \cond B)$ and two of $y: A_2 \cond B$. And so on,
for every nested conditional formula (if $B$ is a conditional
formula $C \cond D$, eight formulas are generated).

A more detailed analysis on the structure of conditionals in the
initial sequent is needed to refine these complexity results. By
the relation between CK+MP and modal logic T, we strongly
conjecture that provability for both BSeqMP and BSeqID+MP is
decidable in $O(n^2logn)$-space. Vigan\`o presents a substructural
analysis for modal system T (chapter 10 of \cite{vigano}) that
could inspire a refinement of the exponential complexity bound we
obtained by the previous informal discussion.

\section{CondLean: A Theorem Prover for Conditional Logics}
In this section we present CondLean, a theorem prover implementing
the sequent calculi SeqS; it is a SICStus Prolog program inspired
by leanTAP \cite{leanTAP}. The program comprises a set of clauses,
each one of them represents a sequent rule or axiom. The proof
search is provided for free by the mere depth-first search
mechanism of Prolog, without any additional ad hoc mechanism.
CondLean is available for free download at
\texttt{http://www.di.unito.it/$\thicksim$olivetti/CONDLEAN.}

We represent each component of a sequent (antecedent and
consequent) by a {\em list} of formulas, partitioned into three
sub-lists: atomic formulas, transitions and complex formulas.
Atomic and complex formulas are represented by a list like
\texttt{[x,a]}, where \texttt{x} is a Prolog constant and
\texttt{a} is a formula. A transition \emph{x $\trans{A}$ y} is
represented by \texttt{[x,a,y]}. For example, the sequent $x: A
\cond B, x: A \cond C, x \trans{A} y \prova y: B, x: C, x: A \imp
B$ is represented by the following lists (the upper one represents
the antecedent, the lower one represents the consequent):
\begin{center}
  \texttt{[[],[[x,a,y]],[[x,a=>b],[x,a=>c]]]}\\
  \texttt{[[y,b],[x,c]],[],[[x,a->b]]}
\end{center}

We present three different implementations:
\begin{enumerate}
  \item a \emph{constant labels} version;
  \item a \emph{free-variables} version;
  \item an heuristic version.
\end{enumerate}
The \emph{constant labels} version makes use of Prolog constants
to represent SeqS's labels. The sequent calculi are implemented by
the predicate
\begin{center}
 \texttt{{\bf prove(Sigma, Delta, Labels)}}.
\end{center}
This predicate succeeds if and only if $\Sigma$ $\prova$ $\Delta$
is derivable in SeqS, where \texttt{Sigma} and \texttt{Delta} are
the lists representing the multisets $\Sigma$ and $\Delta$,
respectively and \texttt{Labels} is the list of labels introduced
in that branch. For example, to prove
\begin{center}
\emph{x: A $\cond$} (\emph{B $\andd$ C})\footnote{CondLean extends
the sequent calculi to formulas containing also $\nott$, $\andd$,
$\orr$ and $\vero$.} $\prova$ \emph{x: A $\cond$ B, x: C}
\end{center}
in CK, one queries CondLean with the goal:
\begin{center}
\texttt{prove([[],[],[[x, a=>(b and c)]]], [[[x,c],[],[[x,
a=>b]]], [x])}.
\end{center}
Each clause of \texttt{prove} implements one axiom or rule of
SeqS; for example, the clause implementing
($\cond$ L) is:\\\\
\begin{footnotesize}
\texttt{ {\bf
prove([LitSigma,TransSigma,ComplexSigma],[LitDelta,TransDelta,\\\indent
\indent
ComplexDelta], Labels)}:-\\
\indent select([X,A=>B],ComplexSigma,ResComplexSigma), member(Y,Labels), \\
\indent put([Y,B],LitSigma,ResComplexSigma,NewLitSigma,NewComplexSigma),\\
\indent prove([LitSigma,TransSigma,ResComplexSigma],\\ \indent
\indent
[LitDelta,[[X,A,Y]$\tc$TransDelta],ComplexDelta],Labels),\\
\indent prove([NewLitSigma,TransSigma,NewComplexSigma],\\ \indent
\indent
[LitDelta,TransDelta,ComplexDelta],Labels).\\\\
}
\end{footnotesize}
The  predicate \texttt{select}
 removes \texttt{[X,A=>B]} from \texttt{ComplexSigma} returning \\
 \texttt{ResComplexSigma} as
result. The predicate \texttt{put} is used to put \texttt{[Y,B]}
in the proper sub-list of the
antecedent.\\
To search a derivation of a sequent $\Sigma$ $\prova$ $\Delta$,
CondLean proceeds as follows. First of all, if $\Sigma$ $\prova$
$\Delta$ is an axiom, the goal will succeed immediately by using
the clauses for the axioms. If it is not, then the first
applicable rule  will be chosen, e.g. if \texttt{ComplexSigma}
contains a formula \texttt{[X,A and B]}, then the clause for
($\andd$ L) rule will be used, invoking \texttt{prove} on the
unique premise of ($\andd$ L). CondLean proceeds in a similar way
for the other rules. The ordering of the clauses is such that the
application of the branching rules is postponed as much as
possible. When the ($\cond$ L) clause is used to prove $\Sigma$
$\prova$ $\Delta$, a backtracking point is introduced by the
choice of a label \texttt{Y} occurring in the two premises of the
rule; in case of failure, Prolog's backtracking tries every
instance of the rule with every available label (if more than
one). Choosing, sooner or later, the right label to apply ($\cond$
L) may strongly affect the theorem prover's efficiency: if there
are \emph{n} labels  to choose for an application of ($\cond$ L)
the computation might succeed only after \emph{n-1} backtracking
steps, with a significant loss of efficiency.

Our second implementation, called {\em free-variables}, makes use
of \emph{Prolog variables} to represent all the labels that can be
used in a single application of the
 ($\cond$ L) rule. This version represents labels by
integers starting from  $1$; by using integers we can easily
express constraints  on the range of the variable-labels. To this
regard the library \texttt{clpfd} is used to manage free-variable
domains (see \cite{CSP} and \cite{CLP} for details about the
constraints satisfaction problems and the constraint logic
programming). As an example, in order to prove $\Sigma^{'}$,
\emph{1: A $\cond$ B} $\prova$ $\Delta$ the theorem prover will
call \texttt{prove} on the following premises: \emph{$\Sigma^{'}$
$\prova$ $\Delta$, 1 $\trans{A}$ V} and \emph{V: B, $\Sigma^{'}$
$\prova$ $\Delta$}, where \emph{V} is a Prolog variable. This
variable will be then instantiated by Prolog's pattern matching to
apply either the (EQ) rule, or to \emph{close a branch with an
axiom}. Here below is the clause implementing the ($\cond$
L) rule:\\\\
\begin{footnotesize}
\texttt{{\bf prove([LitSigma,TransSigma,ComplexSigma],[LitDelta,\\
\indent \indent TransDelta,ComplexDelta],Max)}:-\\ \indent
select([X,A =>
B],ComplexSigma,ResComplexSigma),\\
  \indent domain([Y],1,Max), Y\#>X,\\ \indent
put([Y,B],LitSigma,ResComplexSigma,NewLitSigma,NewComplexSigma),\\
  \indent
  prove([NewLitSigma,TransSigma,NewComplexSigma],\\ \indent \indent
[LitDelta,TransDelta,ComplexDelta],Max),\\
  \indent
  prove([LitSigma,TransSigma,ResComplexSigma],\\ \indent \indent
[LitDelta,[[X,A,Y]$\tc$TransDelta],ComplexDelta],Max).\\\\
 }
\end{footnotesize}
The atom  \texttt{Y\#>X} adds the constraint \texttt{Y>X} to the
constraint
  store: the constraints solver will verify the consistency
  of it during the computation. In SeqCK and SeqID we can only use labels introduced \emph{after}
the label
    \texttt{X}, thus we introduce the previous constraint. In SeqMP and SeqID+MP we can also use
     \texttt{X} itself,
thus we shall add the constraint  \texttt{Y\#>=X}.\\The third
argument of predicate \texttt{prove} is \texttt{Max} and is used
to define variables domains.\\On a sequent with 65 labels on the
antecedent this version succeeds in 460 mseconds, whereas the
constant labels version takes 4326 mseconds.

We have also developed a third version, called {\em heuristic
version}, that performs a "two-phase" computation: in "Phase 1" an
\emph{incomplete}  theorem prover searches a derivation  exploring
a \emph{reduced search space}; in case of failure, the
free-variables version is called ("Phase 2"). Intuitively, the
reduction of the search space in Phase 1 is obtained by committing
the choice of the label to instantiate a free variable, whereby
blocking the backtracking.

For SeqMP and SeqID+MP, we have developed a theorem prover which
simplifies the reformulations given by BSeqMP and BSeqID+MP. In
particular, the reformulation given in the previous section uses
two different auxiliaries sets, namely $K$ and $\Psi$, whereas the
theorem prover implements only $\Psi$, by introducing another
argument \texttt{CondContr} to the predicate \texttt{prove};
therefore, only ($\cond$ L$)_1$ and ($\cond$ L$)_3$ are
implemented. The \texttt{prove} predicate now becomes:
\begin{center}
\texttt{\bf prove(Sigma, Delta, Labels, CondContr)}.
\end{center}
The list \texttt{CondContr} stores the conditional formulas of the
antecedent that have been duplicated so far.
 When ($\cond$ L) is
applied to a formula \emph{x: A $\cond$ B} in the antecedent, the
formula is  duplicated at the same time into the
\texttt{CondContr} list; when ($\cond$ L) is applied to  a formula
in \texttt{CondContr}, in contrast, the formula is \emph{no
longer} duplicated. Thus the ($\cond$ L) rule is split in two
rules, one taking care of "unused" conditionals of the antecedent,
the other taking care of "used" (or duplicated) conditionals.
Observe that this ensures termination. To understand the
difference between the calculus and CondLean implementation, we
can observe that the calculus BSeqMP (BSeqID+MP) ensures that
every conditional formula $x: A \cond B$ is duplicated only once,
no matter how many times it occurs in a branch. As a difference,
CondLean ensures that every \emph{occurrence} of $x: A \cond B$ is
duplicated at most once. However, we have chosen of implementing
this simplified version since it is easier and, at the current
state, it is not clear if the exact implementation of BSeqMP
(BSeqID+MP) would perform significantly better.

\subsection{Performances of the Theorem Prover}
The performances of the three versions of the theorem prover are
promising even on a small machine. To test our program we used
samples generated by modifying the samples from
\cite{free-variable-tableau} and from \cite{vigano}.

We have tested CondLean, SeqCK system,
obtaining the following results\footnote{These results are
obtained running SICStus Prolog 3.10.0 on an Intel Pentium 166
MMX, 96 MB RAM machine.}:
\begin{enumerate}
  \item the constant labels version succeeds in 79 tests over 90
  in less than 2 seconds (78 in less than one second);
  \item the free-variables version succeeds in 73 tests over 90 in less than 2 seconds (but 67 in less
     than 10 mseconds);
  \item the heuristic version succeeds in 78 tests over 90 in less
  than 2 seconds (70 in less than 500 mseconds).
\end{enumerate}

Considering  the sequent-degree (defined as the maximum level of
nesting of the $\cond$ operator) as a parameter, we have the
following results, obtained by testing the SeqCK free-variables
version:
\begin{center} \(\begin{array}{| c | c | c | c | c | c |} \hline
 \mbox{Sequent degree} & 2  &  6  &   9  &  11  &  15 \\
\hline
\mbox{Time to succeed (mseconds)} & 5 &  500  &  650  &  1000  &  2000  \\
\hline
\end{array}\)
\end{center}

We have also tested CondLean, SeqMP system, on some sequents that
require duplications of conditional formulas; in particular, we
have obtained the following results running the heuristic version
on an AMD Athlon XP 2400+ (2.0 GHz), 512 MB RAM machine, using
SICStus Prolog 3.11.1:
\begin{center} \(\begin{array}{| c | c | c | c | c | c | c | c | c | c | c |} \hline
 \mbox{Sequent} & 1  &  2  &   3  &  4  &  5 & 6 & 7 & 8 & 9 & 10 \\
\hline
 \mbox{Number of applications of ($\cond$ L$)_3$} & 1  &  1  &   2  &  2  &  2 & 1 & 2 & 3 & 4 & 5 \\
\hline
\mbox{Time to succeed (mseconds)} & 1 &  2500  &  1  &  1  &  1 & 1 & 1 & 1 & 1 & 1  \\
\hline
\end{array}\)
\end{center}
As expected, in (MP) systems the free-variables version offers
better performances than the constant label version; in the
following table we show how many sequents have been derived by
each implementation in less than 1 ms, 1 second and 2 seconds over
97 valid sequents:
\begin{center} \(\begin{array}{| c | c | c | c |} \hline
 \mbox{Time to succeed} & 1 ms  &  1 s  &   2 s  \\
\hline
\mbox{Constant labels} & 61 &  66  &  67   \\
\hline
\mbox{Free-variables} & 75 &  82  &  82  \\
\hline
\end{array}\)
\end{center}

\section{Conclusions, Comparison with Other Works and Future Work}
In this work we have provided a labelled calculus for minimal
conditional logic CK, and its standard extensions with conditions
ID and MP. The calculus is cut-free and analytic. By a
proof-theoretical analysis, we have shown that CK and CK+ID are
decidable in $O(n\,\log n)$ space. We have also introduced a
decision procedure for CK+MP and CK+MP+ID. To the best of our
knowledge, sequent calculi for these logics have not been
previously studied and the complexity bound for these conditional
systems is new.

We have also developed CondLean, a theorem prover implementing the
calculus written in SICStus Prolog.

We briefly remark on some related works. Most of the works have
concentrated on extensions of CK.

De Swart \cite{7} and Gent \cite{Gent} give sequent/tableaux
calculi for the strong conditional logics VC and VCS. Their proof
systems are based on the entrenchment connective $\leq$, from
which the conditional operator can be defined. Their systems are
analytic  and comprise an infinite set of  rules $\leq F(n,m)$,
with a uniform pattern,  to decompose each sequent with $m$
negative  and $n$ positive entrenchment formulas.

Crocco and Fari{\~{n}}as \cite{CroccoFarinas:95} present  sequent
calculi for some conditional logics including CK, CEM, CO and
others. Their calculi comprise two levels of sequents: principal
sequents with $\vdash_P$ correspond to the basic deduction
relation, whereas auxiliary sequents with  $\vdash_a$ correspond
to the conditional operator: thus the constituents of $\Gamma
\vdash_P \Delta$ are  sequents of the form $X \vdash_a Y$, where
$X,Y$ are sets of formulas.

Artosi, Governatori, and Rotolo \cite{Governatori:02} develop
labelled tableau for the {\em first-degree}  fragment (i.e.
without nested conditionals) of the conditional logic CU that
corresponds to cumulative non-monotonic logics. In their work they
use labels similarly to ours. Formulas are labelled by path of
worlds containing also variable worlds (see also our free-variable
implementation). Differently from us, they do not use a specific
rule to deal with equivalent antecedents of conditionals. They use
instead a  unification procedure to propagate positive
conditionals. The unification process  itself provides to
check  the equivalence of antecedents. Their tableau system is
not analytic, since it contains a cut-rule, called PB, which is
not eliminable. Moreover it is not clear how to extend it to
nested conditionals.

Lamarre \cite{Lamarre:94} presents tableaux systems for the
conditional logics V, VN, VC, and  VW. Lamarre's  method is a
consistency-checking procedure  which tries to build a  system of
sphere falsifying the input formulas. The method makes use of  a
subroutine  to compute the  {\em core}, that is defined as the set
of formulas characterizing the minimal sphere. The computation of
the core   needs in turn the consistency checking procedure. Thus
there is a mutual recursive definition between the procedure for
checking consistency and the procedure to compute the core.

Groeneboer and Delgrande \cite{GroeneboerDelgrande:90} have
developed a tableau method for the conditional logic VN   which is
based on  the translation of this logic into the modal logic S4.3.

\cite{GGOSTableaux2003} have  defined a labelled tableaux
calculus for the logic CE  and some of its extensions. The flat
fragment of  CE corresponds to the nonmonotonic preferential logic
P and admits a semantics in terms of preferential structures
(possible worlds together with  a family of preference
relations). The tableau calculus makes use of pseudo-formulas,
that are modalities in a hybrid language indexed on worlds. In
that paper it is shown how to obtain  a decision procedure for
that logic by performing a kind of loop checking.

Finally, complexity results for some conditional logics
have been obtained by Friedman and Halpern
 \cite{FriedmanHalpern:94}. Their results are based on a
semantic analysis, by an argument about the size of possible
countermodels.

In the future,  we intend to continue  our work in two directions:
\begin{enumerate}
  \item We want to investigate if it is possible to develop sequent calculi
  based on the selection function semantics
  for stronger conditional logics (than CK+MP+ID). If this is possible, we would
  like to extend CondLean to support these stronger systems.

  \item We hope to increase the efficiency of our theorem prover CondLean
  by experimenting standard refinements and heuristics.
\end{enumerate}

\nocite{Governatori:02}
\nocite{free-variable-tableau}
\nocite{leanTAP}
\nocite{cond-e-non-monotonic-2}
\nocite{Boutilier:94}
\nocite{Chellas}
\nocite{counterfactuals}
\nocite{CroccoFarinas:95}
\nocite{crocco2}
\nocite{CroccoLamarre:92}
\nocite{Delgrande:87}
\nocite{cond-e-non-monotonic}
\nocite{GroeneboerDelgrande:90}
\nocite{7}
\nocite{leanTAP-Rev}
\nocite{Fitting}
\nocite{lehmanBel}
\nocite{FriedmanHalpern:94}
\nocite{dov}
\nocite{12}
\nocite{belief-revision}
\nocite{Gent}
\nocite{Ginsberg:86}
\nocite{GiordanoGliozziOlivetti:98}
\nocite{16}
\nocite{GGOSTableaux2003}
\nocite{Grahne:91JLC}
\nocite{Hudelmaier}
\nocite{CLP}
\nocite{KrausLehmannMagidor:90}
\nocite{Lamarre:94}
\nocite{Lamarre:92}
\nocite{Lewis:73}
\nocite{ramsey-test}
\nocite{CSP}
\nocite{Nute80}
\nocite{diagnosi}
\nocite{articoloRoma}
\nocite{articolo-Olivetti}
\nocite{OlivettiSchwind:99}
\nocite{posu}
\nocite{Schwind:99}
\nocite{Stalnaker}
\nocite{vigano}
\nocite{GiordanoSchwind:04}

\bibliographystyle{acmtrans}
\bibliography{articolo}

\end{document}